\documentclass[amssymb,amsmath,aps,secnumarabic,floatfix,rmp,showpacs]{revtex4}
\usepackage{color}
\usepackage{alltt}
\usepackage{psfig}
\usepackage{epsfig}
\usepackage{pstricks,pst-node,pst-tree,pst-grad,graphics}
\usepackage{dcolumn}
\usepackage{xspace}
\usepackage{natbib}
\expandafter\ifx\csname package@font\endcsname\relax\else
 \expandafter\expandafter
 \expandafter\usepackage
 \expandafter\expandafter
 \expandafter{\csname package@font\endcsname}%
\fi

\begin{document}
\title{A new, efficient algorithm for the Forest Fire Model}
\author{Gunnar Pruessner and Henrik Jeldtoft Jensen}
\affiliation{
Department of Mathematics,
Imperial College London,
180 Queen's Gate,
London SW7 2BZ,
UK\\
gunnar.pruessner@physics.org
and
h.jensen@ic.ac.uk
}
\date{\today}
\begin{abstract}
The Drossel-Schwabl Forest Fire Model is one of the best studied models
 of non-conservative self-organised criticality. However, using a new
 algorithm, which allows us to study the model on large statistical and
 spatial scales, it has been shown to lack simple scaling. We thereby
 show that the considered model is not critical. This
 paper presents the algorithm and its parallel implementation in detail,
 together with large scale numerical results for several
 observables. The algorithm can easily be adapted to related problems
 such as percolation.
\end{abstract}
\maketitle
\tableofcontents

\section{Introduction}
The assumption that SOC \citep{Jensen:98} is the correct framework to
describe and explain the ubiquity of power laws in nature, has been
greatly supported by the development of non-conservative models, because
natural processes are typically dissipative. Contrary to these models,
analytical work has suggested, that the deterministic part of the
dynamics must be conservative in order to obtain scale invariance
\citep{Hwa:1989,Grinstein:1990}. However, on a mean-field level, this is
not necessarily true \citep{VespignaniZapperi:1998}, which has been
exemplified in an exact solution of a model, that has a forest fire-like
driving \citep{JensenPruessner:2002a}. However, as a random neighbour
model, the latter lacks spatial extension.

The Drossel-Schwabl Forest Fire Model (DS-FFM)
\citep{DrosselSchwabl:1992} is one of the few spatially extended,
dissipative models, which supposedly exhibit SOC. Contrary to the
Olami-Feder-Christensen stick-slip model
\citep{OlamiFederChristensen:1992}, where criticality is still disputed
(for recent results see for example
\citep{LisePaczuski:2001,LisePaczuski:2001b,BoulterMiller:2003}),
for the DS-FFM the asymptotic divergence of several moments of its statistics, and
therefore the divergence of an upper cutoff can be shown rigorously.
Although this might be considered as a sign of criticality, it is far
from being a sufficient proof. In equilibrium thermodynamics ``criticality''
usually refers to a divergent correlation length
\citep{Binney:98,Stanley:71} in the two-point correlation function, which
is associated with a scale-invariant or power-law like behaviour. This
is how the term ``criticality'' is to be interpreted in SOC: Observables
need to be scale invariant\footnote{In a finite system the
distributions are not expected to be free of any scale, but to be
dominated asymptotically by one scale only.}, i.e. power laws in the
statistics. There are many examples of divergent moments without scale
invariance, such as the over critical branching process
\citep{Harris:1963} or over critical percolation
\citep{StaufferAharonyENG:1994}.

Thus, there is \emph{a priori} no reason to assume that the DS-FFM is
scale free. However, there are many numerical studies, which suggest so
\citep{DrosselSchwabl:1992,Christensen:1993,ClarDrosselSchwabl:1994}, one
of them, however, suggests the breakdown of simple scaling
\citep{Grassberger:1993}. Since an analytical approach is still lacking,
numerical methods are required to investigate this problem. In this
paper, we propose a new, very fast algorithm to simulate the DS-FFM with
large statistics and on large scales. The implementation of the
algorithm, has produced data of very high statistical quality. Some of
the results have been already published elsewhere
\citep{JensenPruessner:2002b}.

The structure of the paper is as follows:
The next section contains the definition of the model together with its
standard observables and their relations. Then the algorithm is
explained in detail. The section finishes with a detailed discussion on
the changes necessary to run the algorithm on parallel or distributed
machines. In the third section results for the two dimensional FFM are
presented and analysed. The paper concludes with a summary in the fourth
section.

\section{Method and Model}
This section is mainly technical: After defining the model, all relevant
details of the implementation are discussed. Apart from concepts such as the
change from a tree oriented algorithm to a cluster oriented algorithm,
concrete technical details are given, for example memory requirements
and methods for handling histograms. The section also contains a
description of the performance analysis of the implementation. A
parallelised version of the algorithm is introduced an discussed in the
last section.

\subsection{The Model} \label{sec:the_model}
A Forest Fire Model was first proposed by Bak, Chen and Tang
\citep{BakChenTang:1990} and changed later by Drossel and Schwabl
\citep{DrosselSchwabl:1992} to what is now known as \emph{the} Forest
Fire Model (or DS-FFM as we call it): On a $d$ dimensional lattice of
linear length $L$, each site has a variable associated with it, which
indicates the state of the site. This can either be ``occupied'' (by a tree),
``burning'' (occupied by a fire) or ``empty'' (ash). In each time step,
all sites are updated in parallel according to the following rules: If a
site is occupied and at least one of its neighbours is burning, it
becomes burning in the next time step. If a site is occupied and none of
its neighbours is burning, it becomes burning with probability $f$. If a
site is empty, it becomes occupied with probability $p$. If a site is
burning it becomes empty in the next time-step with probability
one. As these probabilities become very small, they are better described
as rates in a Poisson like process. From
a simple analysis it is immediately clear
\citep{ClarDrosselSchwabl:1994}, that the model can become critical only in
the limit $p\to 0$ and $f\to 0$. In this limit, the burning process
becomes instantaneous compared to all other processes (see also
sec. \ref{sec:timescales}) and can be represented by the algorithm
shown in \aref{straight}.

\begin{figure}[ht]
\begin{alltt}
FOREVER \{
\emph{   /* Choose a site randomly */}
   rn = random site;
\emph{   /* If empty occupy with probability p */}
   IF (rn empty) THEN \{
      with probability p: rn=occupied;
   \} ELSE \{
\emph{   /* If occupied start a fire with probability f */}
      with probability f: 
         burn entire cluster connected to rn;
   \}
\}
\end{alltt}
\caption{\alabel{straight} The naive, basic algorithm of the DS-FFM}
\end{figure}

Compared to the instantaneous burning, both of the remaining processes
are slow. In section \ref{sec:timescales} it is shown that $p\gg f$ is
required \citep{ClarDrosselSchwabl:1994} for criticality, so that $f/p<1$
and the algorithm in \aref{straight} can be written as
\aref{straight_fast}, which is faster than the former, because
the number of random choices of a site is reduced, but equivalent
otherwise.

\begin{figure}[ht]
\begin{alltt}
FOREVER \{
\emph{   /* The following line is without effect */}
   with probability p: \{ 
      rn = randomly chosen site;
      IF (rn empty) THEN \{
         rn=occupied;
      \} ELSE \{
         with probability f/p: 
            burn entire cluster connected to rn;
      \}
   \}
\}
\end{alltt}
\caption{\alabel{straight_fast} A faster algorithm, doing essentially
 the same as the one shown in \aref{straight}.}
\end{figure}

The line \verb#with probability p# makes sure that the occupation
attempt still happens with probability $p$ and the burning attempt still
occurs with $p f/p = f$. Of course, the line is completely meaningless,
because the alternative, which occurs with probability $1-p$ is no
action at all. It therefore can be omitted. Then every randomly picked
empty site will become occupied, while burning happens with the reduced
probability $f/p$. 

This rescaling of probabilities is only possible in this form if the two
processes are independent, which is the case because a new occupation
can only occur for empty sites, while a burning attempt operates only on
occupied sites. If both processes were to operate on the same type of
site, a reduced probability $(1+f/p)^{-1}$ would decide between the two
alternatives.

The implementation shown in \aref{straight_fast} (without the
meaningless line) has been used for example in
\citep{HoneckerPeschel:1997,Henley:1993}. However, probably for
historical reasons, the model is usually
\citep{Grassberger:1993,ClarDrosselSchwabl:1994,SchenkETAL:2000}
implemented as shown in \aref{straight_traditional}, where trees
are grown in chunks of $p/f$ between two lightning attempts. Although
this means that sites become re-occupied only in chunks of $p/f$,
it turns out that apart from peaks in the histogram of the time series
of global densities of occupied sites \citep{SchenkETAL:2000}, the
statistics do not depend on these details. Only in order to avoid any
confusion, all data for this article have been produced by means of the
algorithm in \aref{straight_traditional}. Moreover this algorithm
is much more suitable for parallelisation (see section
\ref{sec:Parallelizing_the_code}).

\begin{figure}[ht]
\begin{alltt}
FOREVER \{
\emph{   /* This is just a loop to occupy the }
\emph{    * right number of sites */}
   REPEAT p/f TIMES \{
      rn = randomly chosen site;
      IF (rn empty) THEN \{rn=occupied;\}
   \}
   rn = randomly chosen site;
   IF (rn occupied) THEN \{
      burn entire cluster connected to rn;
   \}
\}
\end{alltt}
\caption{\alabel{straight_traditional} The traditional implementation.}
\end{figure}

\subsection{Statistical Quantities} \label{sec:statistical_quantities}
The objects of interest in the DS-FFM are clusters formed by occupied
sites: Two trees belong to the same cluster, if there exists a path
between them along nearest neighbouring, occupied sites. The cluster in
the DS-FFM correspond to avalanches in sandpile-like models
\citep{Jensen:98}. The cluster, which is burnt at each burning step can
be examined more closely, so that various geometrical properties can be
determined either as averages (and higher moments) or as entire
distribution: Mass (in the following this term is used synonymously to
size), diameter, time to burn it etc. The last property is better
expressed as the maximum length for all paths parallel to the axes and
fully within the given cluster, connecting the initially burnt tree and
each tree within the same cluster. It is the maximum number of nearest
neighbour moves one has to make to reach all sites in the same cluster,
in this sense a ``Manhattan distance''
\citep{CormenLeisersonRivest_Manhattan:1990}. As trees catch fire due to
nearest neighbours only, this maximum distance is the total burning time
of the entire cluster. In the definition above, the ``time to burn''
$\manh$ becomes a purely geometrical property of the cluster and
therefore independent from the actual implementation
(see sec.~\ref{sec:burning_procedure}) of the burning procedure.

\subsubsection{Cluster size distribution} \label{sec:clusterdistribution}
The most prominent property of the model, however, is the size
distribution of the clusters, $\dns$, which is the single-site
normalised number density of clusters of mass $s$, i.e. the number of
clusters of size $s$ per unit volume. The average cluster size, i.e. the
average size of a cluster a randomly chosen occupied site belongs to, is
correspondingly defined as
\begin{equation}
 \aves{s} = \frac{\sum_{s} s^2 \dns}{\sum_{s} s \dns} \quad .
\elabel{ave_s}
\end{equation}
As indicated by the bar, $\dns$ denotes the \emph{expected}
distribution, i.e. something to be \emph{estimated} from the
observables. On average, the probability that a randomly chosen site
belongs to a cluster of size $s$ is then $s \dns$. If $n_t(s)$ denotes
the cluster size distribution of the configuration at time $t$ (see
below), then one expects
\begin{equation}
 \ave{n_t(s)} = \dns \quad .
\end{equation}
where $\ave{}$ denotes the ensemble average (as opposed to
$\aves{\ }$, which denotes the average over $s \dnst$). Assuming
ergodicity, one has
\begin{equation}
\lim_{T\to\infty} \frac{1}{T} \sum_{t=1}^T A_t \to \ave{A}
\end{equation}
for an arbitrary quantity $A_t$ measured at each step $t$ of the
simulation. The limit exists for all bound observables $A_t$. 

Regarding the time $t$, it is worth noting that a step in the simulation
is considered completed, i.e. $t \to t+1$ if the randomly chosen site
for the lightning attempt was occupied, i.e. the attempt was successful,
so that $T$ is the number of burnt clusters. For sufficiently large
systems, the changes of the system due to growing or lightning are
almost negligible, and so are the differences between averages taken
over all lightning attempts or all \emph{successful} lightning
attempts. Also, the distributions found directly before and directly
after burning tend to the same expectation value for sufficiently large
systems, see sec.~\ref{sec:finite_size_scaling}. It is noted only for
completeness, that in this paper the cluster size distribution $n_t(s)$
has been measured directly \emph{after} the burning procedure. Therefore
$n_t(s)$ does not include the cluster burnt at time step $t$, just like
$n_{t+1}(s)$ does not in an implementation, where the distribution is
measured \emph{before} burning.

Introducing
\begin{equation} \elabel{def_rho}
 \rhobar = \sum_{s=1} s \dns
\end{equation}
as average density of occupied sites, the expected distribution of
burnt clusters is $s \dns/\rhobar$. To see this, $\PCB_t(s)$ is
introduced, denoting the distribution of clusters burnt in the $t$th
step of the simulation. This distribution contains only one non-zero
value for each $t$, namely $\PCB_t(s)=1$ for the size $s$ of the cluster
burnt at time $t$, and $\PCB_t(s)=0$ for all other $s$. Therefore
\begin{equation}
 \sum_{s=1}^N \PCB_t(s) = 1
\end{equation}
where $N$ is the number of sites in the system, $N=L^d$, which is also
the maximum mass of a cluster.  Since the site where the fire starts is
picked randomly, the cluster burnt in time step $t+1$ is drawn randomly
from the distribution $n_t(s)$ with a probability proportional to the
mass of the cluster. The normalisation of the distribution $s \dns$ is
given by \eref{def_rho}, so that for $t$ large enough, the effect of
the initial condition can be neglected,
\begin{equation} \elabel{converge_pcb}
 \ave{\PCB_t(s)} = s \dns/\rhobar \ .
\end{equation}

In the stationary state the average number of trees, $\rhobar$ is
related to $\aves{s}$ by \citep{ClarDrosselSchwabl:1994}
\begin{equation} \elabel{aves_averho}
 \aves{s} =\frac{1-\rhobar}{\theta \rhobar} \ .
\end{equation}
This equation, as well as \eref{converge_pcb}, is strictly only exact if the density of occupied
sites is constant over the course of the growing phase. For very large
system sizes \eref{aves_averho} holds almost perfectly, as shown in
Tab.~\ref{tab:absolute_results}; however, note the remarks in Sec.~\ref{sec:finite_size_scaling}.

For a coherent picture $\PCA_t(s)$ is introduced, which is the histogram
of \emph{all} clusters, i.e. $\sum_s \PCA_t(s)$ is the number of clusters
in the system at time $t$. According to the definition of $\dns$ it is
\begin{equation} \elabel{converge_pca}
 \ave{\PCA_t(s)} = N \dns \quad ,
\end{equation}
and correspondingly
\begin{equation}
 \rho_t = \frac{1}{N} \sum_s s \PCA_t(s)
\end{equation}
with $\ave{\rho_t}=\rhobar$. Since \eref{converge_pcb} and
\eref{converge_pca} differ on the RHS only by constants rather than by
random variables, both distribution, $\PCB_t(s)$ and $\PCA_t(s)$, are
estimators of the expected distribution $\dns$. Clearly, the burnt
cluster distribution $\PCB_t(s)$ is much sparser than than
$\PCA_t(s)$ and the estimator for $\dns$ derived from this quantity, is
therefore expected to have a significant larger standard deviation. On
the other hand, its autocorrelation time is expected to be considerably
smaller than that of $\PCA_t(s)$, because on average only $p/f+1$
entries ($\rhobar p/f$ sites are occupied in each ``growing loop'', which
is repeated on average $1/\rhobar$ times) of the latter are changed between
two subsequent measurements, corresponding to the number of newly
occupied sites plus the cluster which is burnt down. So, $\PCA_t(s)$
provides a much larger sample size, but is also expected to be much more
correlated. In order to judge, whether it is wise to spend CPU time on
calculating the full $\PCA_t(s)$ rather than only $\PCB_t(s)$, as it was
done in the past \citep{ClarDrosselSchwabl:1994}, these competing effects
need to be considered, by calculating the estimate for the standard
deviation of the estimator of $\dns$ from both observables, which is
discussed in detail in section~\ref{sec:std_details}.

\subsubsection{Timescales} \label{sec:timescales}
In order to obtain critical behaviour in the FFM, a double separation of
time scales is required \citep{ClarDrosselSchwabl:1996}
\begin{equation} \elabel{double_separation}
 f \ll p \ll \left(\frac{f}{p}\right)^{\nu'} \quad ,
\end{equation}
with some positive exponent $\nu'$.
The left relation, $f \ll p$, entails $f/p \to 0$ and therefore
\eref{double_separation} entails $p\to 0$ and $f\to 0$. This is also
the case for
\begin{equation} \elabel{double_separation_sloppy}
 f \ll p \ll 1 \quad ,
\end{equation}
and therefore leads to the same prescription to drive the system,
however \eref{double_separation} entails
\eref{double_separation_sloppy} but not vice versa. This can be seen
by noting that \eref{double_separation} entails the non-trivial
relation $p^{1+1/\nu'} \ll f \ll p$. Some authors, however, just state
\eref{double_separation_sloppy}
\citep{Grassberger:1993,VespignaniZapperi:1998}.  The three scales
involved are due to three different processes and their corresponding
rates: \\
\begin{enumerate}
\item The timescale on which the burning happens, the typical time of
      which is handwavingly estimated as the average number of sites in
      a burnt cluster, $\aves{s} \propto p/f$. A more appropriate
      assumption is that the typical burning time scales like a power of
      the average cluster size \citep{ClarDrosselSchwabl:1996}. This
      should be distinguished from the scaling of the \emph{average}
      time it takes to burn a cluster, because the \emph{typical} time
      represents the chracteristic scale of the burning time
      distribution, which might be very different from its average.\\
\item The timescale of the growing, which is $1/p$. \\ 
\item The timescale of the lightning, $1/f$. \\ 
\end{enumerate}
Burning must be fast compared to growing, so that clusters are burnt
down, before new trees grow on it edges \citep{ClarDrosselSchwabl:1996},
i.e. $(p/f)^{\nu'} \ll 1/p$ or $(f/p)^{\nu'} \gg p$. 
In order to obtain divergent cluster
sizes, growing must be much faster than lightning, i.e. $p \gg f$. Thus,
the double separation reads as stated in \eref{double_separation}.
By making the burning instantaneous compared to all other processes, the
dynamics effectively loses one timescale. In this case, the rates $f$ and $p$,
measured on this microscopic timescale, vanish, i.e. $f=0$ and
$p=0$, so that the right relation of \eref{double_separation} is
perfectly met, provided that $p/f$ does not vanish. However, the ratio $f/p$ remains finite, and $f \ll p$ is
still to be fulfilled. A finite $f/p$ means that one rate provides a scale
for the other. Measuring the rates on the macroscopic timescale, defined
by the sequence of burning attempts, $f$ becomes $1$ in these new
unities, and $p$ becomes $p/f \equiv \theta^{-1}$. The notation
$\theta = f/p$ corresponds to \citep{VespignaniZapperi:1998}, which is,
unfortunately, the inverse of $\theta$ used in
\citep{Grassberger:1993}. \Eref{double_separation} then means
$\theta \to 0$. 
At first sight, this result seems paradoxical, since $\theta=0$
is incompatible with instantenous burning's compliance with $p\ll
\theta^{\nu'}$. However, this problem does not appear in the
\emph{limit} $\theta\to 0$.
In a finite system, one cannot make $\theta$ arbitrarily
small, as the system will asymptotically oscillate between the two states
of being completely filled and completely empty. On the other hand, for
fixed $\theta$ and sufficiently large system sizes, a further increase
in system size will leave the main observables, such as $\rho_t$ and
$\PCA$ (see Section~\ref{sec:clusterdistribution}), essentially
unchanged. These asymptotic values, namely the observables at a given
$\theta$ in the thermodynamic limit, are to be measured.

\subsubsection{Scaling of the cluster size distribution} \label{sec:scaling}
Assuming that finite size effects do not play any r\^ole, i.e. for $\theta$
not too small, the ansatz 
\begin{equation} \elabel{def_tau}
 \dnst = s^{-\tau} \GC(s/\Scutoff(\theta))
\end{equation}
as obtained in percolation \citep{StaufferAharonyENG:1994} is reasonable
for $s$ larger than a fixed lower cutoff. In the following, the
additional parameter $\theta$ in $\dnst$ is omitted, whenever
possible. The quantity $\Scutoff(\theta)$ is the upper cutoff and
supposed to incorporate all $\theta$ dependence of the distribution. It
can be shown easily \citep{ClarDrosselSchwabl:1994} that the second
moment of $\dnst$ (see \eref{ave_s}) diverges in the limit $\theta \to
0$ and $L \to \infty$, so that $\Scutoff$ must diverge with $\theta \to
0$.  Here, $\GC(x)$ plays the r\^ole of a cutoff function, so that
$\lim_{x \to \infty} \GC(x)=0$ and for large $x$ falls off faster than any power,
because all moments of $\dnst$
are finite in a finite system. For finite $x$, $\GC(x)$ can show any
structure and does not have to be constant. However, assuming
$\lim_{\Scutoff \to \infty }\dnst$ finite, $\GC(s/\Scutoff)$ can be
regarded as constant in $s$ for sufficiently large $\Scutoff$, so that $\dnst$
behaves like a power law, $s^{-\tau}$, for certain $s$. However, \emph{a
priori} it is completely unknown, whether $\Scutoff$ is large enough in
that sense and the \emph{only} way to determine $\tau$ directly from
$\dnst$ is via a data
collapse.  It is already known that ``simple scaling'' \eref{def_tau}
does not apply in the presence of finite size effects
\citep{SchenkETAL:2000}.

The assumption \eref{def_tau} states that the FFM is scale-free in the
limit $\Scutoff(\theta) \to \infty$ and \emph{defines} the exponent
$\tau$ which characterises the scale invariance. One cannot stress
enough, that with the breakdown of \eref{def_tau}, the proposed exponent
is undefined, unless a new scaling behaviour is proposed. It has been
pointed out that \eref{def_tau} certainly contains corrections
\citep{Pastor-SatorrasVespignani:2000}. This asymptotic character of the
universal scaling function is well known \citep{Wegner:72} from
equilibrium critical phenomena.

While Grassberger concludes that the ansatz \eref{def_tau} ``cannot
be correct'' \citep{Grassberger:1993}, this is rejected in
\citep{SchenkETAL:2000}. However, the latter authors do not actually investigate
$\GC(x)$ and simply plot their estimate of $s \dnst$
vs. $s/\Scutoff(\theta)$. In the result section it is shown that there
is no reason to believe that \eref{def_tau} could hold in any finite
system.

\subsubsection{Other distributions}
\label{sec:other_dists}
The exponent $\tau$ as defined in \eref{def_tau} can be related to
exponents of other assumed power laws. To this end, the distribution
$\PSF(s, \manh ; \theta)$ is introduced, which is the joint
probability density function (PDF), for a cluster burnt to be of mass
$s$ and burning time (see sec. \ref{sec:statistical_quantities})
$\manh$ at given $\theta$. Then it is possible to define conditional expectation values as
\citep{ChristensenFogedbyJensen:1991}.
\begin{eqnarray}
 {\mathsf E}(s | \manh ; \theta) & = & \sum_{s'} s' \PSF(s', \manh ; \theta) \\
 {\mathsf E}(\manh | s ; \theta) & = & \sum_{\manh'} \manh' \PSF(s, \manh' ; \theta) \quad .
\end{eqnarray}
Moreover it is clear that $\dnst$ is just a marginal distribution, i.e.
\begin{equation}
 s \dnst =  \sum_{\manh'} \PSF(s, \manh' ; \theta) \equiv  \PSF_s(s ; \theta) \quad .
\end{equation}
In the assumed absence of any scale, it is reasonable to define for the
distribution of $\manh$ similar to \eref{def_tau} 
\begin{equation} \label{def:expo_b}
 \PSF_{\manh}(\manh ; \theta) = \manh^{-b} \GC_{\manh}(\manh/\manh_0 (\theta))
\end{equation}
and for the relation between ${\mathsf E}(s | \manh)$ and $\manh$:
\begin{equation} \elabel{Esmanh}
 {\mathsf E}(s | \manh) \propto \manh^{\mu'}
\end{equation}
To avoid confusion, it is important to keep in mind that the absence of
scales is not a physical or mathematical necessity: The system could as
well ``self-organise'' to any other, sufficiently broad distribution,
which could have an intrinsic, finite scale, i.e. a natural constant
characterising the features of the distribution. This looks much less
surprising considering the fact that standard models of critical
phenomena \citep{Stanley:71} like the Ising model, possess such a scale
everywhere apart from the critical point.

An additional assumption is necessary in order to produce a scaling
relation:
\begin{equation} \elabel{asumption_peaked}
 \PSF_{\manh}(\manh ; \theta) d\manh = \PSF_s( {\mathsf E}(s | \manh)  ; \theta) d({\mathsf E}(s | \manh ; \theta)) \quad
\end{equation}
where $\PSF_{\manh}$ and $\PSF_s$ denote the marginal distributions of
$\PSF(s, \manh ; \theta)$, which leads --- assuming sufficiently large
$\Scutoff$ and $\manh_0$ --- to
\begin{equation} \elabel{scaling_relation}
 b = 1 + \mu' (\tau - 2)
\end{equation}
using $\PSF_s=s\dnst$ and \eref{def_tau}.  \Eref{asumption_peaked} is
based on the idea that a cluster requiring burning time $\manh$ is as
likely to occur as a cluster of the size corresponding to the average taken
conditional to the burning time $\manh$. If the distribution
$\PSF(s, \manh ; \theta)$ is very narrow, such that ${\mathsf E}(s | \manh)$ is
virtually the only value of $s$ with non-vanishing
probability\footnote{The extreme case would be $\PSF(s, \manh ; \theta)
= \delta(s-f(\manh)) g(\manh)$ with a monotonic function $f(\manh)$
representing the conditional average.}, this
condition is met. However, the distribution can have any shape and still
obey the assumption, as illustrated in \Fref{peaked_dist}.

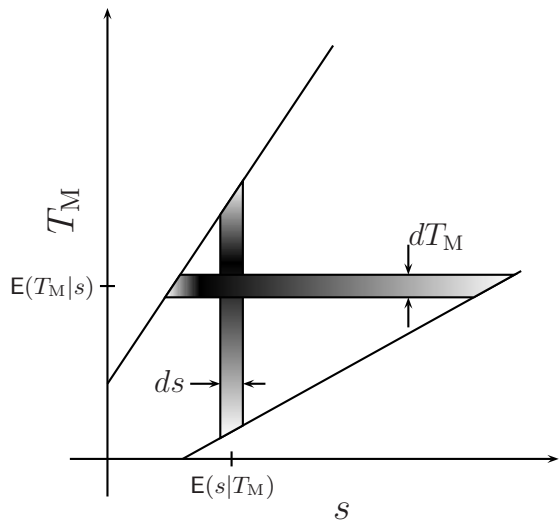
\begin{figure}[th]
\begin{center}
%% 6.5 * 0.8
\begin{pspicture}(0,0)(6.5,6.5)

\psline{->}(0,0.5)(6.5,0.5)
\psline{->}(0.5,0)(0.5,6.5)

\psline(0.5,1.5)(3.5,6.0)
\psline(1.5,0.5)(6.0,3.0)

%%0.5+0.5*(2.5/4.5)
%%1.5+1.5*(4.5/3)
%\psline(2.0,0.77777777)(2.0,3.75)
%%0.5+0.8*(2.5/4.5)
%%1.5+1.8*(4.5/3)
%\psline(2.3,0.94444444)(2.3,4.2)

\pspolygon[fillstyle=gradient,gradbegin=white,gradend=black,gradmidpoint=0.32](2.0,0.77777777)(2.0,3.75)(2.3,4.2)(2.3,0.94444444)
\psline(2.15,0.4)(2.15,0.6)
\put(1.6,0.05){${\mathsf E}(s | \manh)$}

%0.5+1.45*3/4.5
%1.5+2.45*4.5/2.5
%\psline(1.4666,2.95)(5.91,2.95)
%0.5+1.15*3/4.5
%1.5+2.15*4.5/2.5
%\psline(1.2666,2.65)(5.37,2.65)
\pspolygon[fillstyle=gradient,gradbegin=white,gradend=black,gradmidpoint=0.1,gradangle=90](1.4666,2.95)(5.91,2.95)(5.37,2.65)(1.2666,2.65)
\psline(0.4,2.8)(0.6,2.8)
\put(-0.8,2.7){${\mathsf E}(\manh | s)$}

\psline{->}(4.5,3.35)(4.5,2.95)
\psline{->}(4.5,2.25)(4.5,2.65)
\put(4.5,3.35){{\large $dT_{\text{M}}$}}

\psline{->}(1.6,1.5)(2.0,1.5)
\psline{->}(2.6,1.5)(2.3,1.5)
\put(1.12,1.39){{\large $ds$}}

\put(-0.3,3.5){\rotatebox{90}{{\Large $T_{\text{M}}$}}}
\put(3.5,-0.3){{\Large $s$}}

\end{pspicture}
\end{center}
\caption{\flabel{peaked_dist} A schematic joint PDF $\PSF(s, \manh';
\theta )$. The gray shading is used to indicate the density and the
 straight lines indicate roughly the limits of the distribution. While a
 narrower distribution would most easily obey \eref{asumption_peaked},
 it does not necessarily have to be sharply peaked. In this example the
 weighted areas of the horizontal and the vertical stripes might be the
 same. They cross at the conditional averages.}
\end{figure}

Scaling relation \eref{scaling_relation} can only be derived via
\eref{asumption_peaked}, which cannot be mathematically correct, as
$\PSF_s$ is actually only defined for integer arguments, while in general
${\mathsf E}(s | \manh)$ is not integer valued. However, the
scaling relation might hold in some limit. 

The exponent defining the divergence of $\Scutoff$ in \eref{def_tau} is
defined as
\begin{equation} \elabel{def_scutoff}
 \Scutoff(\theta) = \theta^{-\lambda}
\end{equation}
leading together with \eref{ave_s} and \eref{aves_averho}
to the scaling relation \citep{ClarDrosselSchwabl:1996}
\begin{equation}
 \lambda (3-\tau) =1 \quad .
\elabel{scaling_relation_ltau}
\end{equation}
The corresponding exponent for $\manh_0$ in (\ref{def:expo_b}) as
\begin{equation}
 \manh_0(\theta) = \theta^{-\nu'}
\elabel{def_nu}
\end{equation}
The assumption $\manh_0 = {\mathsf E}(\manh | \Scutoff) \propto \Scutoff^{1/\mu'}$
then gives the scaling relation 
\begin{equation}\elabel{scaling_relation_nup}
 \nu' = \frac{\lambda}{\mu'}
\end{equation}
It is interesting to note that this assumption is consistent with the
assumption that clusters, which have a size of the order
$\Scutoff(\theta)$ need of the order $\manh_0$ time to burn. In
that case one has $\PSF_\manh(\manh_0;\theta) d\manh =
\PSF_s(\Scutoff;\theta) ds$ and as $\manh_0 \propto
\Scutoff^{\nu'/\lambda}$, one has using (\ref{def:expo_b}) and
\eref{def_tau}: 
\begin{equation}
 (1-b)\frac{\nu'}{\lambda}=2-\tau
\end{equation}
corresponding to \eref{scaling_relation} with \eref{scaling_relation_nup}.

\subsection{The Implementation} \label{sec:implementation}
In this section the new implementation of the DS-FFM is discussed. An
implementation especially capable to handle large scales has been
proposed by Honecker \citep{HoneckersFFMCode} earlier. The most prominent
feature of it is the bitwise encoding of the model, which significantly
reduces memory requirements. Some of the properties investigated, profit
from this scheme of bitwise encoding, because bitwise logical operators
can be used to determine for example correlations, and operate on entire
words ``in parallel''. However, in this implementation it would have
been inefficient to count all clusters, i.e. $\dns$ is determined via
$\PCB(s)$ rather than $\PCA(s)$.

In contrast to standard implementations
\citep{ClarDrosselSchwabl:1994,SchenkETAL:2000,HoneckerPeschel:1997},
where $\dns$ is derived from $\PCB(s)$, the philosophy of the
implementation presented in this article is to count \emph{all} clusters
efficiently by keeping track of their growing and disappearance, so that
$\dns$ is derived from $\PCA(s)$. By comparing the standard deviation of
the estimates, and the costs (CPU time), the efficiency is found to be
at least one order of magnitude higher. At the same time, the complexity
of the algorithm is essentially unchanged, namely $\OC(\INFL \log(N))$
instead of $\OC(\INFL)$, while a naive implementation of the counting of
all clusters is typically of order $\OC(N)$. In the following the
algorithm is described in detail. Because of its close relation to
standard percolation, the algorithm presented below is also applicable
for this classical problem of statistical mechanics. In fact, the
percolation algorithm recently proposed by Newman and Ziff
\citep{NewmanZiff:2000,NewmanZiff:2001} is very similar. Based on many
principles presented in this paper, an asynchronously parallelised
version for percolation has been developed recently \citep{MoloneyPruessner:2003}.

\subsubsection{Tracking clusters}
\label{sec:tracking_clusters} Usually each site is represented by a
two-state variable, which indicates whether the site is occupied or
empty. The variable does not need to indicate the state ``burning'',
because the burning procedure is instantaneous compared to all other
processes and can be implemented without introducing a third state
(see sec.~\ref{sec:burning_procedure}). In order to keep track of the cluster
distribution, each site gets associated two further variables (in an
actual implementation the number of variables can be reduced to one,
see sec.~\ref{sec:reducing_memory}), one which
points (depending on the programming language either directly as an
address or as an index) to its ``representative'' and one which contains
the mass of the cluster the given site is connected to. The
representative of a site is another site of the same cluster, but not
necessarily and in fact typically not a nearest neighbour. This is shown
in \fref{lattice_figure}. If a site is empty, the pointer to a
representative is meaningless. The pointer of representatives form a
tree-like structure, because representatives might point to another
representative, as shown in \fref{tree_figure}. A site which
points to itself and is therefore its own representative, is called a
``root'' site, since it forms the root of the tree like structure. Only
at a root site, the second variable, denoting the mass of the cluster, is
actually meaningful and indicates the mass of the entire cluster. Each
cluster is therefore uniquely identified by its root site: Any two
sites, which belong to the same cluster have the same root and vice
versa. By construction of the clusters (shown below), it takes less than
$\OC(\log N)$ to find the root of any site in the system.

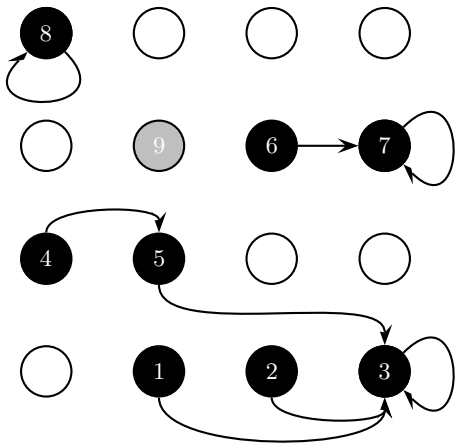
\begin{figure}[th]
\begin{center}
%% 6.5 * 0.8
\begin{pspicture}(0,0)(6.5,6.5)
\psset{radius=.23}
\psset{xunit=1cm,yunit=1cm}
\psset{arrowsize=0.2}
\newcommand{\mynode}[2]{ \rput(#1){\circlenode{#2}{ {\color{white} #2 }}} }

\mynode{1.0,1.0}{0}
\psset{fillstyle=solid,fillcolor=black}
\mynode{2.5,1.0}{1}
\mynode{4.0,1.0}{2}
\mynode{5.5,1.0}{3}
\psset{fillstyle=solid,fillcolor=white}

\psset{fillstyle=solid,fillcolor=black}
\mynode{1.0,2.5}{4}
\mynode{2.5,2.5}{5}
\psset{fillstyle=solid,fillcolor=white}
\mynode{4.0,2.5}{0}
\mynode{5.5,2.5}{0}

\mynode{1.0,4.0}{0}
\psset{fillstyle=solid,fillcolor=lightgray}
\mynode{2.5,4.0}{9}
\psset{fillstyle=solid,fillcolor=black}
\mynode{4.0,4.0}{6}
\mynode{5.5,4.0}{7}
\psset{fillstyle=solid,fillcolor=white}

\psset{fillstyle=solid,fillcolor=black}
\mynode{1.0,5.5}{8}
\psset{fillstyle=solid,fillcolor=white}
\mynode{2.5,5.5}{0}
\mynode{4.0,5.5}{0}
\mynode{5.5,5.5}{0}

\nccurve[ncurv=0.5,angleA=270,angleB=270]{->}{1}{3}
\nccurve[ncurv=0.5,angleA=270,angleB=270]{->}{2}{3}
\nccurve[ncurv=0.5,angleA=270,angleB=90 ]{->}{5}{3}
\nccurve[ncurv=5,angleA=45,   angleB=315]{->}{3}{3}
%%\nccurve[ncurv=0.5,angleA=180,angleB=180]{->}{5}{6}
\nccurve[ncurv=0.5,angleA=0,  angleB=180]{->}{6}{7}
\nccurve[ncurv=5,angleA=45,   angleB=315]{->}{7}{7}
\nccurve[ncurv=5,angleA=315,  angleB=225]{->}{8}{8}
\nccurve[ncurv=0.5,angleA=90, angleB=90 ]{->}{4}{5}
\psset{fillstyle=solid,fillcolor=black}
\mynode{1.0,5.5}{8}
\mynode{5.5,4.0}{7}
\mynode{5.5,1.0}{3}
\psset{fillstyle=solid,fillcolor=white}
%%\nccircle{->}{B4}{0.5}
%% \nccurve[ncurv=5,angleA=270,  angleB=270]{->}{B4}{B4}
\end{pspicture}
\end{center}
\caption{All occupied sites (black) on the lattice point to a representative. The
 site pointing to itself is the root of the cluster. The site shown in
 light gray is the one which is about to become occupied, as shown in
 \fref{cluster_join}. The labels on the sites are just to uniquely
 identify them in other figures.
\flabel{lattice_figure}}
\end{figure}

\begin{figure}[th]
\begin{center}
%%% 6.5 * 0.8
%\psset{radius=.23}
%\psset{arrows=<-}
%\psset{fillstyle=solid,fillcolor=black}
%\newcommand{\mynode}[1]{ \Tcircle{ {\color{white} #1 }} }
%\pstree[treemode=U]{\mynode{3}}
%{
%\mynode{1}
%\mynode{2}
%\pstree[treemode=U]{\mynode{5}}{\mynode{4}}
%}

%% 6.5 * 0.8
\begin{pspicture}(0,0)(5.0,5.5)
\psset{radius=.23}
\psset{xunit=1cm,yunit=1cm}
\psset{arrowsize=0.2}
\newcommand{\mynode}[2]{ \rput(#1){\circlenode{#2}{ {\color{white} #2 }}} }
\psset{fillstyle=solid,fillcolor=black}

%\mynode{4.5,5.0}{4}
%\mynode{0.5,3.0}{1}
%\mynode{2.5,3.0}{2}
%\mynode{4.5,3.0}{5}
%\mynode{2.5,1.0}{3}

\mynode{3.5,5.0}{4}
\mynode{0.5,3.0}{1}
\mynode{2.0,3.0}{2}
\mynode{3.5,3.0}{5}
\mynode{2.0,1.0}{3}

\psset{fillstyle=solid,fillcolor=white}

\nccurve[ncurv=0.5,angleA=270,angleB=90]{->}{4}{5}
\nccurve[ncurv=0.5,angleA=270,angleB=90]{->}{2}{3}
\nccurve[ncurv=0.5,angleA=315,angleB=135]{->}{1}{3}
\nccurve[ncurv=0.5,angleA=225,angleB=45]{->}{5}{3}

\nccurve[ncurv=5,angleA=315,  angleB=225]{->}{3}{3}
\psset{fillstyle=solid,fillcolor=black}
\mynode{2.0,1.0}{3}
\end{pspicture}
\end{center}
\caption{The tree-like structure of the largest cluster in \fref{lattice_figure}.
\flabel{tree_figure}}
\end{figure}

The algorithm is a dynamically updated form of the Hoshen-Kopelman
algorithm \citep{HoshenKopelman:1976}. The same technique has recently
been used to simulate percolation efficiently for many different
occupations densities \citep{NewmanZiff:2000}. The method described in
the following differs from \citep{NewmanZiff:2000}, by not only growing
clusters, but also removing them. While one of the main advantages of
the original Hoshen-Kopelman algorithm is its strong reduction of memory
requirements to $\OC(L^{d-1})$, the algorithm described here
only makes use of the data representation proposed by Hoshen and
Kopelman, so that the memory requirements are still $\OC(L^d)$.

In the following the technique, how to create and to update the clusters,
is described in detail. 

Starting from an empty lattice, the first site becomes occupied by
setting the state variable. Since this site cannot be member of a larger
cluster, its representative is the site itself. Therefore the mass
variable must be set to one. The same pattern applies to all other sites
which get occupied, as long as they are isolated. The procedure becomes
more involved, when a site induces a merging of clusters. This is the
case whenever one or more neighbours of the newly occupied site are
already occupied. In general the procedure is then as follows:
\begin{itemize}
\item Find the root of all neighbouring clusters.
\item Reject all roots, which appear more than once in
      order to avoid double counting.
\item Identify the largest neighbouring cluster. 
\item Increase the mass variable of the root of this cluster by the mass
      of all remaining clusters (ignoring those which have been rejected
      above) plus one (for the newly occupied site).
\item Bend the representative pointers of the roots of all remaining
      clusters to point to the root of the largest cluster (keeps the
      tree height small, see below).
\item Bend the representative pointers of the newly occupied site to
      point to the root of the largest cluster.
\end{itemize}

\begin{figure}[th]
\begin{center}
%% 6.5 * 0.8
\begin{pspicture}(0,0)(6.5,6.5)
\psset{radius=.23}
\psset{xunit=1cm,yunit=1cm}
\psset{arrowsize=0.2}
\newcommand{\mynode}[2]{ \rput(#1){\circlenode{#2}{ {\color{white} #2 }}} }

\mynode{1.0,1.0}{0}
\psset{fillstyle=solid,fillcolor=black}
\mynode{2.5,1.0}{1}
\mynode{4.0,1.0}{2}
\mynode{5.5,1.0}{3}
\psset{fillstyle=solid,fillcolor=white}

\psset{fillstyle=solid,fillcolor=black}
\mynode{1.0,2.5}{4}
\mynode{2.5,2.5}{5}
\psset{fillstyle=solid,fillcolor=white}
\mynode{4.0,2.5}{0}
\mynode{5.5,2.5}{0}

\mynode{1.0,4.0}{0}
\psset{fillstyle=solid,fillcolor=darkgray}
\mynode{2.5,4.0}{9}
\psset{fillstyle=solid,fillcolor=black}
\mynode{4.0,4.0}{6}
\psset{fillstyle=solid,fillcolor=darkgray}
\mynode{5.5,4.0}{7}
\psset{fillstyle=solid,fillcolor=white}

\psset{fillstyle=solid,fillcolor=black}
\mynode{1.0,5.5}{8}
\psset{fillstyle=solid,fillcolor=white}
\mynode{2.5,5.5}{0}
\mynode{4.0,5.5}{0}
\mynode{5.5,5.5}{0}

\nccurve[ncurv=0.5,angleA=270,angleB=270]{->}{1}{3}
\nccurve[ncurv=0.5,angleA=270,angleB=270]{->}{2}{3}
\nccurve[ncurv=0.5,angleA=270,angleB=155 ]{->}{5}{3}
%%\nccurve[ncurv=0.5,angleA=180,angleB=180]{->}{5}{6}
%%\nccurve[ncurv=0.5,angleA=270, angleB=90]{->}{6}{3}
\nccurve[ncurv=0.5,angleA=0,  angleB=180]{->}{6}{7}
\nccurve[ncurv=0.5,angleA=0,   angleB=0 ]{->}{7}{3}
\nccurve[ncurv=5,angleA=315,  angleB=225]{->}{8}{8}
\nccurve[ncurv=0.5,angleA=90, angleB=90 ]{->}{4}{5}
\nccurve[ncurv=0.5,angleA=290,angleB=155 ]{->}{9}{3}
\psset{fillstyle=solid,fillcolor=black}
\mynode{1.0,5.5}{8}
\psset{fillstyle=solid,fillcolor=white}
\mynode{4.0,2.5}{0}
%%\nccircle{->}{B4}{0.5}
%% \nccurve[ncurv=5,angleA=270,  angleB=270]{->}{B4}{B4}
\end{pspicture}
\end{center}
\caption{The configuration in \fref{lattice_figure} after
 occupying the highlighted site. Sites, the pointer of which have been
 changed, are shown in dark gray (site $6$, $7$ and $9$).\flabel{cluster_join}}
\end{figure}
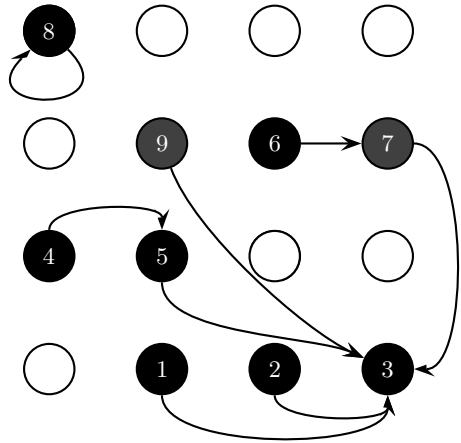

This procedure is depicted in \fref{cluster_join}, illustrating the join
of the clusters shown in \fref{lattice_figure}. As an optimisation, one
could also bend the pointer of site $6$ to point to site $3$, which
would effectively be a form of path compression. However, as shown
below, the trees generated have only logarithmic height, so that the
path compression possibly costs more CPU time than it saves for system
sizes reachable with current computers\footnote{Similarly for other
forms of path compression, for example bending the pointer of the
preceeding to the adjacent site in \texttt{find\_root} (\aref{find_root}).}. It is
important to note that only the root of the largest cluster is not
redirected.

\begin{figure}[ht]
\begin{alltt}
\emph{/* Find the root of the cluster identified by start_index. }
\emph{ * All sites are expected to have a pointer to their }
\emph{ * representative in the array pointer_of. The result}
\emph{ * is stored in index. */}
index = start_index
WHILE ( index != pointer_of[index] ) \{
  index=pointer_of[index]; \}
\end{alltt}
\caption{\alabel{find_root} The \texttt{find\_root} algorithm. All sites are
 expected to have a pointer to their representative in the array
 \texttt{pointer\_of}. The result of this procedure is
 \texttt{index}.}
\end{figure}

To find the root of a given site, which is necessary, whenever clusters
are considered for merging, an algorithm like the one shown in
\aref{find_root} needs $\OC(h_m(M(\CC)))$ time (worst case), where
$h_m(M(\CC))$ is the maximum height of a cluster containing $M(\CC)$
sites, $\CC$ being the cluster under consideration.

All clusters are constructed by merging clusters, which might often involve
single sites. These clusters are represented as trees, like the one
shown in \Fref{tree_figure}. In the following this representation is
used. By construction, if at least two trees join, the resulting tree
has either the height of the tree representing the largest cluster or the
height of any of the smaller trees plus one --- whatever is
larger. Thus, by construction,
\begin{equation} \elabel{hmhm}
 h_m(M) \ge h_m(M') \text{ for any } M \ge M' \quad ,
\end{equation}
so in order to find the maximum height of a tree of mass $M$, one has to
consider the worst case when the smaller trees have maximum height. For
a given, fixed $M$, this is the case when only two cluster merge, so
\begin{equation}
 h_m(M) \le \max\Big( \max_{M' \le \lfloor \frac{M}{2} \rfloor} (h_m(M-M')), \max_{M' \le \lfloor \frac{M}{2} \rfloor} ( 1+h_m(M')) \Big) \quad ,
\end{equation}
where $\lfloor \frac{M}{2} \rfloor$ denotes the integer part of $M/2
\ge 0$, which is is the maximum size of the smaller cluster. The outer
$\max$ picks the maximum of the two $\max$ running over all allowed sizes
of the smaller cluster.
Using \eref{hmhm},
\begin{equation}
 h_m(M) \le \max(h_m(M-1), 1+h_m(\lfloor \frac{M}{2} \rfloor))
\end{equation}
so that
\begin{equation}
 h_m(M) \le \left\{ 
\begin{array}{lr}
1 + h_m(\lfloor \frac{M}{2} \rfloor) & \text{ for } 1 + h_m(\lfloor \frac{M}{2} \rfloor) \ge h_m(M-1) \\
& \\
h_m(M-1) & \text{ otherwise }
\end{array}
\right.
\end{equation}
With $h_m(1)=1$ one can see immediately that
\begin{equation}
 h_m(M) \le \lceil \log_2(M) \rceil \quad 
\end{equation}
by induction, nothing that $\lceil \log_2(M/2) \rceil = \lceil \log_2(M)
\rceil -1$, where $\lceil a \rceil \equiv \lfloor a \rfloor + 1$ for any
$a\ge 0$. Hence 
\begin{equation}
\elabel{complexity_find}
 h_m(M) \in \OC(\log(M)) \quad ,
\end{equation} 
which is therefore the (worst case) complexity of the algorithm. It is
worthwhile noting that all the algorithms considered are just one
solution of the more general union-find (and also insert) problem
\citep{CormenLeisersonRivest:1990}.

As the tree constructed is directed, there is no simple way to find all
sites which are pointing to a given site. This means that splitting
trees is extremely expensive in terms of complexity. However, in the
DS-FFM trees do not get removed individually, but always as complete
clusters. Thus, no part of the tree structure needs to be updated during
the burning (s. section \ref{sec:burning_procedure}).

\subsubsection{Reducing memory requirements} \label{sec:reducing_memory} 
The three variables (state, pointer, size) mentioned above would require
a huge amount of memory: At least a bit for the state (but for
convenience a byte), a word for the address and a word for the mass
(actually depending on the maximum size of the clusters). However, as
the pointers are only meaningful if the site is occupied, the
representative pointer can also be used to indicate the state of a site:
If it is $0$ (or \verb#NULL# if it is an address), the site is empty and
occupied otherwise.

As mentioned above (section \ref{sec:tracking_clusters}), the mass
variable is meaningful only at a root site.  Since only a certain range
of pointers is meaningful, the remaining range can be used to indicate
the mass of a cluster. Assuming that indeces can only be positive,
negative numbers as the value of the pointer can be interpreted as
self references and their modulus as total mass of the cluster. The
concept is restricted to system sizes which are small enough that the
space not occupied by meaningful pointers is large enough to store the
mass information. How large is the maximum representable system size
(not to be confused with memory requirements, which is $N$ times
word size)?  For a word size of $b=4$ byte, i.e. $M=2^{8b}$
representable values in a word, the maximum system size is $N=2^{31}-1$,
namely $-1 \dots -N$ values for indicating masses, $1 \dots N$ for
indeces and $0$ for the empty site, summing up to $2N+1 \le M$, which is
overruled by the memory required $b N \le M$, as $M$ is (usually) the
maximal addressable memory for a single process.
 
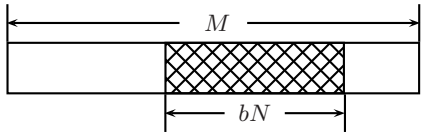
\begin{figure}[th]
\begin{center}
%% 6.5 * 0.8
\begin{pspicture}(0,0)(6.5,2.5)
%%\psset{radius=.23}
%%\psset{xunit=1cm,yunit=1cm}
%%\psset{arrowsize=0.2}
%%\psset{linewidth=0.8pt}
%%\newcommand{\mynode}[2]{ \Cnode(#1){#2} }
\psframe[fillstyle=solid,fillcolor=white](0.5,1.0)(6.0,1.7)
\psframe[fillstyle=crosshatch,linecolor=black](2.6,1.0)(5.0,1.7)

\rput(0.4pt,0){\psline[](2.6,1.0)(2.6,0.5)}
\psline[](5.0,1.0)(5.0,0.5)

\psline[]{<-}(2.6,0.75)(3.4,0.75)
\psline[]{<-}(5.0,0.75)(4.2,0.75)
\rput(3.8,0.75){$b N$}

\rput(0.4pt,0){\psline[](0.5,1.7)(0.5,2.2)}
\rput(-0.4pt,0){\psline[](6.0,1.7)(6.0,2.2)}

\psline[]{<-}(0.5,1.95)(2.8,1.95)
\psline[]{<-}(6.0,1.95)(3.7,1.95)
\rput(3.3,1.95){$M$}

\end{pspicture}
\end{center}
\caption{The memory layout when using addresses as pointers to
 representative. The hatched area is used for valid addresses, what
 remains left can be used to represent cluster masses, i.e. if the value
 of an address points into the white area, the value is interpreted as a
 mass. \flabel{continuous_chunk}}
\end{figure}

When using addresses as pointers, it is less obvious how to identify the
range of meaningless pointers which could be used to store the mass
information.  In order to distinguish quickly whether a given value is
an address or a mass, the most obvious way is to use higher bits in the
pointers. What is the range of meaningless addresses? The addresses are
words, occupying $b N$ bytes. If each byte is individually addressable
(as usual), their value differs by at $b$, i.e. they span a range of $b
N$ different values. As shown in \fref{continuous_chunk}, the
largest remaining continuous chunk of values, not used for references to
representatives, has therefore at least size $\lceil (M-b N)/2 \rceil = (M-b
N)/2$, assuming that the pointer values used, which is also the range of
addresses where they are stored, spans a continuous range.
If the $N+1$
different cluster masses are to be represented as pointer values
pointing into the meaningless region, one has $1+N \le (M - b N)/2 $, i.e. $(b+2)N + 2 \le
M$. If they do not have to be continuous, the condition is relaxed:
$1 + N \le M - bN$. Alternatively one can make use of the lower bits: If
the pointers point to words in a continuous chunk of memory or at least
are all aligned in the same way, then all pointers are identical
$\pmod{b}$, i.e. all pointers $p$ obey $p = c \pmod{b}$ where $0\le c <b$ is a 
constant. Since $b>1$ one can use $p \ne c \pmod{b}$ to
indicate that a given pointer value is to be interpreted as mass, which
can easily be calculated via a bit-shift.

In C it is reasonable to represent the sites as \verb#void *# and
interprete these as pointers to other sites, i.e. \verb#void **#, so that
the loop to search for a root just becomes the code shown in
\aref{Cfind_root}.

\begin{figure}
\begin{alltt}
void *start_pointer, *root, *content;
\emph{/* start_pointer is the address of the site, the root of which}
\emph{ * is to be found. root will always point to the site currently}
\emph{ * under consideration, while content is always the address}
\emph{ * root is pointing to. }
\emph{ * The macro IS_SIZE verifies, whether the value given is a size. */}


\emph{/* Initialise: Assume that start_pointer is the root and }
\emph{ * read its content. */}
for (content=*((void **)(root=start_pointer));

\emph{/* Test whether root's content is actually a size. */}
    (!IS_SIZE(content));

\emph{/* Iterate: content is not a size, so the next candidate}
\emph{ * is what root is currently pointing to. }
\emph{ * Content is updated accordingly. */}
    content=*((void **)(root=content)));
\end{alltt}
\caption{\alabel{Cfind_root} An implementation of
 \texttt{find\_root} in C using pointers to \texttt{void}.}
\end{figure}

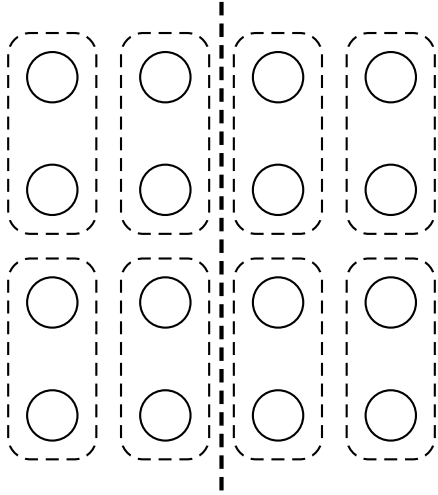
\begin{figure}[th]
\begin{center}
\rotatebox{90}{%% 6.5 * 0.8
\begin{pspicture}(0,0)(6.5,6.5)
\psset{radius=.23}
\psset{xunit=1cm,yunit=1cm}
\psset{arrowsize=0.2}
\newcommand{\mynode}[2]{ \rput(#1){\circlenode{#2}{ {\color{white} #2 }}} }

\psframe[linestyle=dashed,fillstyle=solid,fillcolor=white,framearc=0.5](0.4,0.4)(3.1,1.6)
\psframe[linestyle=dashed,fillstyle=solid,fillcolor=white,framearc=0.5](3.4,0.4)(6.1,1.6)

\psframe[linestyle=dashed,fillstyle=solid,fillcolor=white,framearc=0.5](0.4,1.9)(3.1,3.1)
\psframe[linestyle=dashed,fillstyle=solid,fillcolor=white,framearc=0.5](3.4,1.9)(6.1,3.1)

\psline[linestyle=dashed,linewidth=1.6pt](0,3.25)(6.5,3.25)

\psframe[linestyle=dashed,fillstyle=solid,fillcolor=white,framearc=0.5](0.4,3.4)(3.1,4.6)
\psframe[linestyle=dashed,fillstyle=solid,fillcolor=white,framearc=0.5](3.4,3.4)(6.1,4.6)

\psframe[linestyle=dashed,fillstyle=solid,fillcolor=white,framearc=0.5](0.4,4.9)(3.1,6.1)
\psframe[linestyle=dashed,fillstyle=solid,fillcolor=white,framearc=0.5](3.4,4.9)(6.1,6.1)

\mynode{1.0,1.0}{0}
\mynode{2.5,1.0}{1}
\mynode{4.0,1.0}{2}
\mynode{5.5,1.0}{3}

\mynode{1.0,2.5}{4}
\mynode{2.5,2.5}{5}
\mynode{4.0,2.5}{0}
\mynode{5.5,2.5}{0}

\mynode{1.0,4.0}{0}
\mynode{2.5,4.0}{9}
\mynode{4.0,4.0}{6}
\mynode{5.5,4.0}{7}

\mynode{1.0,5.5}{8}
\mynode{2.5,5.5}{0}
\mynode{4.0,5.5}{0}
\mynode{5.5,5.5}{0}
\end{pspicture}}
\end{center}
\caption{If occupied, each site within a dashed box belongs to the
 same cluster. On a triangular lattice the dashed patches would be
 triangular, each one containing three sites. The thick dashed line
 shows the orientation of the boundary between two consecutive slices
 in the parallelised code, see Sec.~\ref{sec:Parallelizing_the_code}.
\flabel{subgroupify}}
\end{figure}

Representing each site by a word instead of a byte or even a bit
\citep{HoneckersFFMCode}, still leads to reasonably small memory
requirements for typical system sizes (for instance a system of size
$N=4096 \times 4096$ would require $64\MB$). Since the
algorithm has an almost random memory access pattern, it is not
reasonable to implement it out of core \citep{DowdSeverance:1998}. In
order to simulate even larger sizes, the following representation has
been implemented: At the beginning of the simulation the entire lattice
is splitted in cells so that whatever site in such a cell is occupied,
it must belong to the same cluster as any other occupied site in the
same cell, i.e. each site in the cell is
nearest neighbor of all other sites in the cell. On an hyper-cubic lattice these cells
have size $2$, as depicted in \fref{subgroupify}: Each site
within such a cell must belong to the same cluster if it is
occupied. Therefore only one pointer is necessary to refer to its
representative. On a triangular lattice these cells would have size
$3$. Since a pointer can be non-null, although not all sites in the cell
are occupied, a new variable must represent the state of the sites in
each cell, if not lower or higher order bits of the pointers can be used
(see above). On the hyper-cubic lattice the memory requirement is therefore
for each pair of sites $2$ bit for the state and $1$ word for the
address or index of the representative. Storing the $2$ bits in a byte
(and keeping the remaining $6$ bits unused), the memory requirements are
therefore reduced to $(b + 1)N/2$ bytes. Using indices the maximum
representable system size is given by $3/2 N +1 \le M$ and using
pointers with a size identification as shown in \Fref{continuous_chunk} the
constraint is $1+N \le (M-\frac{bN}{2})/2 $ in worst case.

\subsubsection{Efficient histogram superposition}
\label{sec:efficient_histogram_superposition} So far, only the
maintenance of the cluster structure has been described. Since the
masses of all clusters involved are known, it is simple to maintain a
histogram of the cluster mass distribution: If a cluster of size $s$ is
burnt, the corresponding entry in $\PCA_t(s)$ is decreased by one. If a
cluster changes size, $\PCA_t(s)$ is updated accordingly. For example,
when two clusters of size $s_1$ and $s_2$ merge as a particular site is
newly occupied during the growing procedure, $\PCA_t(s_1)$ and
$\PCA_t(s_2)$ are decreased by one and $\PCA_t(s_1+s_2+1)$ is increased
by one. 

Na\"{\i}vely, the average cluster size distribution is the average of
$\PCA_t(s)$, i.e.
\begin{equation}
  \frac{1}{T}\sum_{t'=1}^t  \PCA_{t'}(s) 
\end{equation} 
with $T$ as the number of iterations. Depending on the resolution of the
histogram, it would be very time consuming to calculate this sum for
each $s$. Using exponential binning (which is in fact a form of hashing)
in order to reduce the size of the histogram solves the problem only
partly.

Ignoring any hashing, a na\"{\i}ve superposition, where each slot in
the histogram needs to be read, has complexity $\OC(T H)$, where $H$ is
the largest cluster mass in the histogram.

This problem is solved by noting that early changes in the histogram
propagate though the entire sequence of histograms. Denoting the initial
histogram as $\PCA_0(s)$ and $\Delta \PCA_t(s) = \PCA_{t-1}(s) - \PCA_t(s)$
then
\begin{equation}
 \PCA_t(s) = \PCA_0(s) + \sum_{t'=1}^t \Delta \PCA_{t'}(s)
\end{equation}
and therefore
\begin{equation} \elabel{histo_superposition}
 \sum_{t=1}^T \PCA_t(s) = T \PCA_0(s) + \sum_{t'=1}^T (T-t'+1) \Delta \PCA_{t'}(s) \quad .
\end{equation}
By using this identity only the right hand side of
\eref{histo_superposition} is maintained by increasing it by $T-t+1$
when a new cluster is created at time $t$ and by decreasing it by the
same amount when it is destroyed. In this way, the complexity is only of
order $\OC(T (\INFL +1))$, according to the number of clusters created
and destroyed, i.e. the number of changes in the distribution. This
concept becomes only problematic, if floating point numbers are used to
store the histogram and the accuracy is so small that changes in the sum
by $1$ do not change the result anymore.\footnote{For integers the
precision is not a problem, but the maximum representable number easily
becomes a problem.} The maximum value in $\PCA_t(s)$, where this does
not happen, is given by the largest $m$ with $m + 1 \ne m$ where $m$ is
a variable of the same type as $\PCA_t(s)$. For floating point number,
the value of $m$ is related to the constant
\verb#DBL_EPSILON# (or \verb#FLT_EPSILON# for single precision), which
essentially characterises the length of the mantissa. The concrete value
of $m$ is actually platform, precision and type dependent. For an
unsigned integer of size $4$, this value would be $(2^{32}-1)-1$,
corresponding to $\verb#ULONG_MAX# - 1$, for double precision IEEE75
floating point numbers this value is $\verb#FLT_RADIX**DBL_MANT_DIG#
-1$, i.e. $2^{53}-1$.

Provided that the right hand side of \eref{histo_superposition} is below
the threshold $m$ discussed above for all $s$, this means that only a single
histogram needs to be maintained. It is initialised with $T \PCA_0(s)$
and updated with $\pm(T-t+1)$ at time step $t$, when a cluster of size
$s$ appears or disappears. It is worth mentioning that this concept
obviously even works in conjunction with binning (or any other hashing).

\subsubsection{Implementation of the burning procedure} \label{sec:burning_procedure}
The burning procedure was implemented in the obvious way, without making
use of the tree structure, as shown in \aref{burning}. Although
the burning procedure could also be implemented explicitly recursively,
it is of course significantly faster when implemented iteratively. The
usage of a stack in the procedure might be thought of as reminiscent of the
underlying recursive structure.
\begin{figure}[ht]
\begin{alltt}
\emph{/* Initialise current_stack. */}
CLEAR current_stack;
\emph{/* Put initial site on current_stack. */}
PUT rn ON current_stack;
\emph{/* Sites are cleared right after they have entered the current_stack. */}
rn = empty;
\emph{/* The first loop runs until there is nothing left to}
\emph{ * burn, i.e. next_stack was not filled during the inner loop. */}
DO \{
\emph{   /*  Clear next_stack so that it can get filled in the next loop. */}
   CLEAR next_stack;
\emph{   /* The next loop runs as long as there are sites left to burn}
\emph{    * in the current generation of the fire. */}
   WHILE current_stack not empty \{
\emph{      /* GET: remove the upmost element from current_stack and}
\emph{       * put it in x */}
      GET x FROM current_stack;
\emph{      /* Visit all neighbours */}
      FOR all neighbours n of x \{
         if (n occupied) \{
\emph{            /* Put occupied sites on the current_stack of the next}
\emph{             * generation of the fire */}
            PUT n ON next_stack;
            n = empty
         \}
      \}
   \}
\emph{/*  The next current_stack to be considered is next_stack. */}
   current_stack = next_stack;
\} WHILE current_stack is not empty
\end{alltt}
\caption{\alabel{burning} The burning procedure starting at
 \texttt{rn}. In an actual implementation the copying of
 \texttt{next\_stack} to \texttt{current\_stack} can easily be omitted by
 repeating the code above with \texttt{current\_stack} and \texttt{next\_stack}
 interchanged, similar to a red-black approach
 \citep{DowdSeverance:1998}.}
\end{figure}
The number of times the outer loop in \aref{burning} runs,
defines the generation of the fire front and gives $\manh$; other
properties of the burnt cluster can be extracted accordingly.  The most
important resource required by this procedure are the stacks: One for
the currently burning sites and one for the sites to be burnt in the
next step. There is no upper limit known for the number of
simultaneously burning sites, apart from the naive $N/2$ on a hyper-cubic
lattice, which comes from the observation that sites, which belong to
the same generation of the fire, must reside on the same sub-lattice
(even or odd).

On the other hand it is also trivial to find the maximal number of sites
which burn at the same time, if the fire starts in a completely dense
forest, i.e. in a lattice with $\rho=1$. Obviously the size of the $t$th
generation is then given by $4(t-1)$ for $t>1$ and $1$ at the beginning,
$t=1$. Since the sum of these numbers is the number of sites which is
reachable within a certain time $t$, \emph{the sum} is also an upper
limit for the number of simultaneously burning sites. Indeed, the actual number
can easily be larger than $4(t-1)$, caused by arrangements of wholes
in the lattice, which delay the fire spreading at certain sites, so that
they burn together with a larger fire front. Such a construction is shown
in \fref{delayed_burning}.

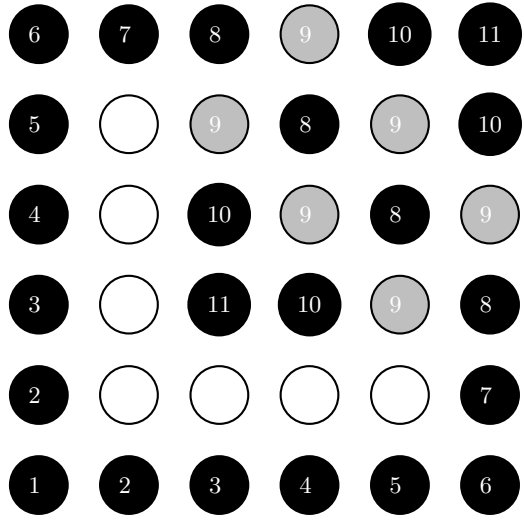
\begin{figure}[th]
\begin{center}
%% 6.5 * 0.8
\begin{pspicture}(0,0)(7.6,7.6)
\psset{radius=.23}
\psset{xunit=0.8cm,yunit=0.8cm}
\psset{arrowsize=0.2}
\newcommand{\mynodeb}[2]{ \psset{fillstyle=solid,fillcolor=black} \rput(#1){\circlenode{#2}{ {\color{white} #2 }}} \psset{fillstyle=solid,fillcolor=white}}
\newcommand{\mynodew}[2]{ \psset{fillstyle=solid,fillcolor=white} \rput(#1){\circlenode{#2}{ {\color{white} #2 }}} \psset{fillstyle=solid,fillcolor=white}}
\newcommand{\mynodeg}[2]{ \psset{fillstyle=solid,fillcolor=lightgray} \rput(#1){\circlenode{#2}{ {\color{white} #2 }}} \psset{fillstyle=solid,fillcolor=white}}
\newcommand{\myspace}{ }

\mynodeb{1.0,1.0}{1\ }
\mynodeb{2.5,1.0}{2\ }
\mynodeb{4.0,1.0}{3\ }
\mynodeb{5.5,1.0}{4\ }
\mynodeb{7.0,1.0}{5\ }
\mynodeb{8.5,1.0}{6\ }

\mynodeb{1.0,2.5}{2\ }
\mynodew{2.5,2.5}{3\ }
\mynodew{4.0,2.5}{4\ }
\mynodew{5.5,2.5}{5\ }
\mynodew{7.0,2.5}{6\ }
\mynodeb{8.5,2.5}{7\ }

\mynodeb{1.0,4.0}{3\ }
\mynodew{2.5,4.0}{4\ }
\mynodeb{4.0,4.0}{11}
\mynodeb{5.5,4.0}{10}
\mynodeg{7.0,4.0}{9\ }
\mynodeb{8.5,4.0}{8\ }

\mynodeb{1.0,5.5}{4\ }
\mynodew{2.5,5.5}{5\ }
\mynodeb{4.0,5.5}{10}
\mynodeg{5.5,5.5}{9\ }
\mynodeb{7.0,5.5}{8\ }
\mynodeg{8.5,5.5}{9\ }

\mynodeb{1.0,7.0}{5\ }
\mynodew{2.5,7.0}{6\ }
\mynodeg{4.0,7.0}{9\ }
\mynodeb{5.5,7.0}{8\ }
\mynodeg{7.0,7.0}{9\ }
\mynodeb{8.5,7.0}{10}

\mynodeb{1.0,8.5}{6\ }
\mynodeb{2.5,8.5}{7\ }
\mynodeb{4.0,8.5}{8\ }
\mynodeg{5.5,8.5}{9\ }
\mynodeb{7.0,8.5}{10}
\mynodeb{8.5,8.5}{11}

\end{pspicture}
\end{center}
\caption{The burning order for a $6\times 6$ patch of sites, where seven
 sites are not occupied and form a barrier, such that some sites behind
 it burn later, together with the fire front propagating away from the
 starting point of the fire at the lower left hand corner. The sites
 belonging to the largest set of trees burning at the same time are
 shown in light gray, unoccupied sites are shown in white, occupied
 sites in black. The numbers indicate the generation of the fire, which
 is one plus the Manhattan distance from the starting point of the fire along
 occupied sites.\flabel{delayed_burning}}
\end{figure}

Of course it is neither reasonable, nor practically possible to provide
enough memory for the theoretical worst case, i.e. two stacks each of
size $N/2$. Indeed the typical memory requirements seem to be of order
$\OC(\sqrt{\INFL})$, as shown in Table~\ref{tab:performance_data}, where
$\maxff$ denotes the largest fire front observed during the simulation.
Providing stacks only of size $4 L$ turned out to be a failsafe, yet
pragmatic solution. Formally one could implement a slow out-of-core
algorithm in the rare yet possible case the memory for the stack is insufficient, i.e. use
hard-disk space to maintain it. In fact, this is what \emph{de facto}
happens if one uses a stack of size $N/2$ on a virtual memory system.

\begin{table}
\caption{\label{tab:performance_data}
Performance data for different parameters and setups.
``ap3000,2'' denotes a parallel run on two nodes on an AP3000,
accordingly ``ap3000,4''. ``cluster,10'' denotes a cluster of 25 Intel
machines, connected via an old 10 MBit network, ``cluster,100'' denotes
the same cluster on a 100 MBit network. ``single1'' and ``single2''
denote two different types of single nodes.
The largest fire front, $\maxff$, was only measured on these
systems. The quantity $\cpuratio$ is the ratio of the average time
(real time on the parallel systems in order to include communication overhead, user time on single nodes) 
for one successful update during
statistics, i.e. when all data structures need to be maintained, and
 equilibration (transient) i.e. when the standard representation is used.}
\newcolumntype{d}[1]{D{.}{.}{#1}}
\begin{tabular}{l|r|r|d{2}|d{3}}
System & L & $\INFL$ & \cpuratio & \maxff \\
\hline
ap3000,2    & 8000   & 4000       & 1.51         & \\
ap3000,2    & 8000   & 8000       & 1.52         & \\
\hline
ap3000,4    & 16000  & 4000       & 1.34         & \\
ap3000,4    & 16000  & 8000       & 1.48         & \\
ap3000,4    & 16000  & 16000      & 1.37         & \\
ap3000,4    & 16000  & 32000      & 1.41         & \\
\hline
cluster,10  & 32000  & 4000       & 2.71         & \\
cluster,10  & 32000  & 64000      & 3.81         & \\
\hline
cluster,100 & 32000  & 32000      & 1.76         & \\
\hline
single1     & 1000   & 500        & 1.41         & 216 \\
single1     & 2000   & 1000       & 1.41         & 326 \\
single1     & 4000   & 125        & 1.42         & 106 \\
single1     & 4000   & 250        & 1.47         & 172 \\
single1     & 4000   & 500        & 1.48         & 255 \\
single1     & 4000   & 1000       & 1.53         & 317 \\
single1     & 4000   & 2000       & 1.50         & 518 \\
single1     & 4000   & 4000       & 1.57         & 646 \\
single1     & 4000   & 8000       & 1.48         & 907 \\
single1     & 4000   & 16000      & 1.45         & 1327 \\
\hline
single2     & 8000   & 4000       & 2.11         & 687 \\
single2     & 8000   & 8000       & 2.11         & 912 \\
single2     & 8000   & 16000      & 2.09         & 1415 \\
\end{tabular}
\end{table}

\subsubsection{Complexity of the algorithm}
The overall complexity of the algorithm has two contributions: The
``growing'' part, where new clusters are generated from existing ones
and the ``burning'' part. The time needed for the burning part is
proportional to the number of sites burnt and therefore expected as
$\OC(\LS)$ (see \eref{ave_s} and \eref{aves_averho}) and $\OC(N)$ in the worst case. 
Since $\aves{\rho}$ in \eref{aves_averho} is bound, the complexity
of ``burning'' is $\OC(\INFL)$ (expected). The complexity of 
``growing'' is estimated by the average number of sites newly occupied,
$\INFL$, times the worst-case complexity \eref{complexity_find} to
find the root of any given site, because up to four roots need to be
found at each tree growing. According to \eref{complexity_find} the
worst case complexity to find the root of any given site is
$\OC(\log(N))$, leading to an overall complexity for ``growing'' of
$\OC(\log(N) \INFL) \supset \OC(\INFL)$. In practice the logarithmic correction
is negligible, especially since $\log(N)$ is an extreme overestimate of
the average case and therefore essentially the same runtime-behaviour is
expected for both procedures \citep{NewmanZiff:2001}. 

Implementations like the one in \citep{HoneckersFFMCode} avoid this logarithmic
factor by counting only the burnt cluster and therefore arrive at an
overall complexity of $\OC(\INFL)$.

The algorithm presented has therefore only a negligibly higher
computational complexity compared to implementations which measure only
$\PCB$. This is corroborated by the comparison of the CPU time per
burnt cluster during equilibration, i.e. the transient, when the cluster
structure do not need to be maintained and the algorithm used is the
standard implementation, to the CPU time per burnt cluster during
statistics, i.e. when observables are actually measured and especially
$\PCA$ is produced. This ratio is shown as $\cpuratio$ in
Tab.~\ref{tab:performance_data} and Tab.~\ref{tab:corrtimes}. It varies only
slightly with $L$ or $\INFL$.

Apparently the algorithm presented offers more statistics, however it
suffers from one limitation: It requires about $(b + 1)N/2$ bytes memory
(see section \ref{sec:reducing_memory}), compared to $N/8$ bytes in
bitwise implementations like \citep{HoneckersFFMCode}, i.e. typically a
factor $20$ more. In order to ascertain whether this disadvantage is
acceptable with respect to the statistical gain, one has to determine
the standard deviations of the calculated quantities for both
implementations.

\subsection{Calculating the standard deviation} \label{sec:std_details}
In order to compare the two algorithm rigorously, it is necessary to
estimate the standard deviation of the estimators for $\dns$ produced by
them \citep{MuellerBinder:73,LandauBinder:2000}:
\begin{equation} \elabel{stds}
\begin{array}{rcl}
 \sigma^2_{\PCB}(s) & = & \frac{2\tau_{\PCB} + 1}{T-1} 
  \Big(\big\langle \PCB_t(s)^2 \big\rangle - \big\langle \PCB_t(s) \big\rangle^2\Big) \\
&&\\
 \sigma^2_{\PCA}(s) & = & \frac{2\tau_{\PCA} + 1}{T-1}  
  \Big(\big\langle \PCA_t(s)^2 \big\rangle - \big\langle \PCA_t(s) \big\rangle^2\Big)
\end{array}
\end{equation}
Here $\tau_{\PCB}$ and $\tau_{\PCA}$ are the correlation times of the
two quantities. Calculating the correlation time in the standard fashion
by recording the history $\PCA_t(s)$ and $\PCB_t(s)$ for each $s$ would
mean to store millions of floating point numbers. Therefore it was
decided to restrict these calculations to just a small yet representative
set of $s$ values. The result shows that the standard deviation does not
fluctuate strongly in $s$.

Because of the special form of $\PCB_t(s) \in {0,1}$, its variance is
particularly simple,
\begin{equation}
 \big\langle \PCB_t(s)^2 \big\rangle = \big\langle \PCB_t(s) \big\rangle
\end{equation}
so that
\begin{equation}
 \sigma^2_{\PCB}(s) = \frac{2\tau_{\PCB} + 1}{T-1} \big\langle \PCB_t(s) \big\rangle \Big(1-\big\langle \PCB_t(s) \big\rangle\Big) \quad .
\end{equation}

The correlation time of $\PCB_t(s)$ is expected to be extremely small,
not only on physical grounds --- an cluster can only burn down once ---
but also because of the extreme dilution of $\PCB_t(s)$, as it was
described in section~\ref{sec:clusterdistribution}. For fixed $s$, most of the
$\PCB_t(s)$ are $0$. In contrast, the $\PCA_t(s)$ are expected to have a
large correlation time, because ``only'' $\INFL+1$ entries are changed
between two subsequent histograms.

The correlation function is calculated in the symmetric way as proposed in
\citep{Anderson:71}, here for an arbritrary quantity $A_t$:
\begin{equation}
 \phi_{t'}^{AA} = 
\frac{\langle A_t A_{t+t'} \rangle_{T-t'} - 
\langle A_t \rangle_{T-t'} \langle A_{t+t'} \rangle_{T-t'}}
{\langle A_t^2 \rangle_{T} -
\langle A_t \rangle_{T}^2}
\end{equation}
where $\ave{}_{T-t'}$ denotes the average taken over time $t$ from $t=1$
to $t=T-t'$. The quantity $\phi_{t'}^{AA}$ was fitted to
$\exp(-t/\tau_A)$ in order to find the correlation time $\tau_A$. The
results are given in Table~\ref{tab:corrtimes}.

\begin{table}
\caption{\label{tab:corrtimes}
Correlation times $\tau_b$ and $\tau_a$ of the corresponding observables
 $\PCB$ and $\PCA$ as a function of $s$ and for different parameters
 $L$, $\INFL$. Values of $s$ marked by ``B'' are results for bins around
 the $s$ value indicated.  
For each set of parameters, the quantity $\cpuratio$ is
 given. It denotes the ratio between the average CPU-time
 for one successful update during equilibration (transient) and during
 statistics, see also Tab.~\ref{tab:performance_data}. The two fractions
$\frac{\sqrt{\sigma^2_{\PCB}(s)}}{\langle \PCB_t(s) \rangle}$, 
$\frac{\sqrt{\sigma^2_{\PCA}(s)}}{\langle \PCA_t(s) \rangle}$, 
their
 ratio $\alpharatio$
and
 $\alpharatio^2/\cpuratio$ are derived.
$\ast$ marks cases, wheres $\tau_b(s)=0$ has been assumed. $\est$ marks
 values of $\tau_a(s)$, which have been extrapolated from $\tau_a(s)$
 for smaller $s$. 
}
\newcolumntype{d}[1]{D{.}{.}{#1}}

\begin{tabular}{r|r|d{2}|r|d{3}|d{2}|d{5}|d{5}|d{1} | d{1}}
$L$ & 
$\INFL$ & 
\cpuratio & 
$s$ & 
\multicolumn{1}{c|}{$\tau_b(s)$} & 
\multicolumn{1}{c|}{$\tau_a(s)$} & 
\multicolumn{1}{c|}{$\frac{\sqrt{\sigma^2_{\PCB}(s)}}{\langle \PCB_t(s) \rangle}$} &
\multicolumn{1}{c|}{$\frac{\sqrt{\sigma^2_{\PCA}(s)}}{\langle \PCA_t(s) \rangle}$} &
\multicolumn{1}{c|}{$\alpharatio$} &
\multicolumn{1}{c}{$\alpharatio^2/\cpuratio$} \\
\hline
4000 & 4000  & 1.57 & $10$     &  \noth  &  \noth &   0.0138\ast &  \noth    &  \noth &  \noth  \\
&&&                   $100$    &  0.170  &   23.6 &   0.0637     & 0.00099   &  64.3  &  2633.4 \\
&&&                  B $10^3$  &  0.028  &   14.2 &   0.0450     & 0.00191   &  23.6  &   354.8 \\
&&&                  B $10^4$  &  0.006  &   10.0 &   0.0412     & 0.00470   &   8.8  &    49.3 \\
&&&                  B $10^5$  &  \noth  &    7.2 &   0.0662\ast & 0.02104   &   3.1  &     6.1 \\
\hline                                                    
4000 & 16000 & 1.45 & $10$     &  0.013  &   39.9 &   0.0141     & 0.00056   &  25.4  &  444.9 \\
&&&                   $100$    &  0.126  &   28.8 &   0.0608     & 0.00127   &  48.0  & 1589.0 \\
&&&                  B $10^3$  &  0.006  &    4.7 &   0.0457     & 0.00175   &  26.1  &  469.0 \\
&&&                  B $10^4$  &  0.013  &    2.9 &   0.0512     & 0.00332   &  15.4  &  163.6 \\
&&&                  B $10^5$  &  \noth  &    2.2 &   0.0433     & 0.00795   &   5.4  &   20.1 \\
\hline                                                    
8000 & 1000  & \noth& $10$     &  0.131  &   \noth&   0.0154     &  \noth    &  \noth & \noth \\
&&&                   $100$    &  0.122  &  284.6 &   0.0602     & 0.00158   &  38.1  & \noth \\
&&&                  B $10^3$  &  0.028  &  236.5 &   0.0399     & 0.00337   &  11.8  & \noth \\
&&&                  B $10^4$  &  0.016  &  163.5 &   0.0397     & 0.00878   &   4.5  & \noth \\
\hline                                                    
8000 & 4000  & 2.11 & $10$     &  0.122  &   78.2 &   0.0154     & 0.00052   &  29.8  &  420.9 \\
&&&                   $100$    &  0.132  &   16.4 &   0.0634     & 0.00087   &  72.9  & 2518.7 \\
&&&                  B $10^3$  &  0.022  &    8.2 &   0.0438     & 0.00147   &  29.7  &  418.1 \\
&&&                  B $10^4$  &  0.005  &    5.5 &   0.0442     & 0.00241   &  18.3  &  158.7 \\
&&&                  B $10^5$  &  \noth  &    4.2 &   0.0409\ast & 0.01006   &   4.1  &    8.0 \\
&&&           B $2\cdot 10^5$  &  \noth  &    3.8\est &  0.0635\ast & 0.02055   &   3.1 &  4.6 \\
\hline                                                    
8000 & 16000 & 2.09 & $10$     &  \noth  &  262.5 &   0.0139\ast & 0.00068  &  20.5   &  201.1 \\
&&&                   $100$    &  0.131  &   56.1 &   0.0629     & 0.00087  &  72.0   & 2480.4 \\
&&&                  B $10^3$  &  0.014  &   19.0 &   0.0467     & 0.00115  &  40.6   &  788.7 \\
&&&                  B $10^4$  &  0.009  &   11.1 &   0.0503     & 0.00296  &  17.0   &  138.3 \\
&&&                  B $10^5$  &  0.006  &    8.3 &   0.0411     & 0.00689  &   6.0   &   17.2 \\
&&&           B $2\cdot 10^5$  &  \noth  &    7.5 &   0.0423\ast & 0.00947  &   4.5   &    9.7 \\
&&&           B $5\cdot 10^5$  &  \noth  &    7.0\est&   1.1106\ast & 0.33331  &   3.3 &   5.2 
\end{tabular}
\end{table}

As described in \eref{converge_pcb} and \eref{converge_pca}, the
two estimators for $\dns$ differ slightly. However, except for $\dns$ only constant
values appear on the RHS of \eref{converge_pcb} and
\eref{converge_pca}, so that the relative errors of $\langle
\PCB_t(s) \rangle_T$ and $\langle \PCA_t(s) \rangle_T$ are also the
relative errors of the estimators for $\dns$ derived from them. These
relative errors are shown in Tab.~\ref{tab:corrtimes} as well. Their ratio is
given as $\alpharatio$ and is an indicator for the advantage of the
algorithm proposed. If the relative error is to be improved by a factor
$q$, one needs to invest $q^2$ CPU-time, i.e. if the algorithm proposed
in this paper costs a factor $\cpuratio$ more CPU-time, and the gain in
the relative error $\alpharatio$, the total gain is
$\alpharatio^2/\cpuratio$. The values for this quantity are also given
in table~\ref{tab:corrtimes}.

According to the table, for fixed $\theta$ relative errors and the
correlation times are only weakly affected by an increase in system size. At
first sight, this is counter-intuitive, as the number of passes
\citep{Henley:1993,HoneckerPeschel:1997}, the mean number of times a site
has been visited between two lightnings, decreases inversely proportional
to the total number of sites in the system: $1/(\theta \rhobar L^2
)$, see Section~\ref{sec:tree_density}. Assuming that this number is
essentially responsible for the error, suggests to keep the number of
passes constant among different $L$. However, this is apparently not the
case, possibly because of self-averaging \citep{FeLaBi:91} effects.

The table also shows various tendencies, which are worth
mentioning. First of all, the total gain becomes smaller for larger
avalanche size $s$. The B in front of some of the values indicates that
a bin around the $s$ value was investigated, 
i.e. the time series of
\begin{equation}
\sum_{s' \in \BC} \PCAB(s')
\end{equation}
was considered, where $\BC$ is a set of (consecutive) $s$ values, representing the
bin. For larger values of $s$, these sets get exponentially larger,
which is necessary for a reasonably large number of events as basis for
the estimators. The general tendency that the proposed algorithm is even
more efficient at small $s$ is not surprising: $\PCB$ samples from $s
\dns$, while $\PCA$ samples only from $\dns$, i.e. $\PCB$ ``sees''
larger cluster more often. \emph{Nevertheless $\PCA$ still is
advantageous by roughly a factor $5$.} The empty entries in
Tab.~\ref{tab:corrtimes} are due to numerical inaccuracies or simply
missing simulations for certain parameters. Some entries are estimated
and marked as such.

There is an additional correlation not mentioned so far: The individual
points in the estimator of the distribution $\PCA$ are not
independent. There are ``horizontal correlations'', i.e. $\PCA(s)$ is
correlated for different values of $s$. These are additional
correlations due to clusters of small sizes, which are likely to grow
and propagate through $s$ in $\PCA_t(s)$ for consecutive time
steps, i.e.
\begin{equation}
\ave{\PCA_t(s) \PCA_{t'}(s')} - \ave{\PCA_t(s)} \ave{\PCA_{t'}(s')}  \quad .
\end{equation}
This correlation is at least partly captured by the correlations
measured for the binned data. It is to be distinguished from the
correlations of \emph{independent} realisations, where correlations are
expected in the cluster size distribution also, i.e.
\begin{equation}
 \ave{\PCA_t(s) \PCA_t(s')} - \ave{\PCA_t(s)} \ave{\PCA_t(s')} \quad .
\end{equation}

This must be taken into account as soon as estimates of $\dns$ for
different $s$ are compared, as it is done when an exponent is calculated
by fitting. This effect is also present for $\PCB$, which is, however,
diluted so enormously that it influences the outcome only in an
insignificant way.

The horizontal correlations could be estimated using a Jackknife scheme
\citep{Efron:82}, similar to that used to calculate the error bar of the
exponent from the time evolution of a quenched Ising model
\citep{SchLoiPru:2001}. While it is certainly essential for the careful
estimation of the error bar of an exponent, it is irrelevant for the
discussion in this paper, as it is quantitatively based only on
\emph{local} comparisons of error bars (overlaps), while its global
properties, i.e. shape and collapse with other histograms estimated, is
not concerned with errors bars. Some authors even seem to dismiss the
relevance of these correlations completely \citep{NewmanZiff:2001}.

\subsection{Parallelising the code}
\label{sec:Parallelizing_the_code} Constructing clusters and keeping
track of clusters rather than of single sites seems to be in
contradiction to any attempt to run the algorithm distributed, that is
splitting the lattice into $S$ {\it slices} (one-dimensional
decomposition --- as periodic boundaries apply, the slices may better be
called cylinders). Moreover, there is a general problem of
parallelisation which becomes apparent in this context: The usual
bottleneck of parallel systems is the communication layer. In order to
keep the communication between sub-lattices as low as possible, fast
parallel code on a lattice requires as few interaction between slices as
possible, while the whole point of doing physics on large lattices is
the assumption of significant interaction between their parts. It is
this fundamental competition of requirement and basic assumption which
makes successful parallel code so rare and which seems to indicate that
problems must have very specific characteristics in order to be
parallelisable in a reasonable way.

However, it is indeed possible to run the algorithm described above on
parallel machines successfully in the sense that it does not only make
use of the larger amount of (distributed) memory available, but also of
the larger amount of computing capabilities. In fact, the code was
successfully rewritten using MPI \citep{GroppLuskSkjellum:1999} and has
been run on two systems with distributed memory: The massively parallel
machine AP3000 at the Department of Computing at Imperial College and on
a cluster of workstations (25 nodes).

In the following the most important design characteristics are described
which proved important in order to make the code running reasonably
fast. This concerns mainly the statistics part, but the equilibration
also needs some tricks. 

MPI assures that packets sent from one node to another in a certain
order are received in exactly the same order --- in the language of MPI
this means that the message ordering is preserved in each particular
communicator. But, how different communicators relate to each other,
i.e. how one stream of packets relates to another one is not specified. If,
for instance, node A sends a packet to node B, and then to node C, which
then sends a packet to node B, this packet might arrive earlier at B
then the packet first sent by A, see \fref{mpi_message_order}.

\begin{figure}[th]
\begin{center}
\begin{pspicture}(0,0)(3.5,3.5)
\psset{xunit=.70710cm,yunit=1cm}
\psset{arrowsize=0.2}
\newcommand{\mynode}[2]{ \rput(#1){\circlenode{#2}{ {\color{white} #2 }}} }

\rput(0.5,0.5){\circlenode{A}{A}}
\rput(4.5,0.5){\circlenode{B}{B}}
\rput(2.5,2.5){\circlenode{C}{C}}
\ncline{->}{A}{B}\naput{1}
\ncline{->}{A}{C}\naput{2}
\ncline{->}{C}{B}\naput{3}

\end{pspicture}
\end{center}
\caption{\flabel{mpi_message_order} 
Nodes $A$, $B$ and $C$ send messages in the order indicated. However, it might
 well happen that the message sent last by node $C$ to node $B$, namely
 message $3$, arrives at that node before message $1$, sent
 \emph{before} message $2$ was sent, which arrived \emph{before} message
 $3$ was sent.}
\end{figure}
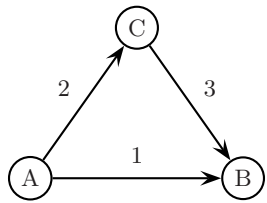

However, it is one of the main goals of parallelisation, to avoid any
kind of synchronisation, which is extremely expensive. Even in a
master-slave design, as it was chosen here, one encourages communication
between the slaves, whenever they can anticipate what to do next or can
indicate each other what to do next.

As explained above (\ref{sec:the_model}) an update consists essentially
of two steps: Growing and burning. Both processes now are distributed
among the slices. The growing procedure is realized by trying to grow
$\INFL/S$ trees in each slice. This is not an exact representation of a
growing procedure taking place on the entire lattice at once, because
the latter has a non-vanishing probability to grow all trees at one
particular spot, while the parallelised version distributes them evenly
among the different slices. Provided that $\INFL$ is large compared to
$S$, this effect can certainly be neglected. The advantage of the
procedure is that the growing procedure at each slice does not need to
be conducted by the master. The burning procedure is more complex, as
the fire starts at one particular site of the entire lattice, so that it
must be selected by the master. The exact procedure of the possibly
following burning process depends on the stage of the algorithm.

In the following the procedures are explained in terms of ``sites''
rather than ``cells'', as introduced in section
\ref{sec:reducing_memory}. Using cells instead of sites makes the code
slightly more complicated, but the changes are obvious. If the cells are
oriented parallel to the borders of slices (see
\fref{subgroupify}), so that its width is a multiple of $2$ in
case of a hyper-cubic lattice, the algorithm runs considerably faster, as
the communication between the nodes is reduced by the same factor. 

\subsubsection{Equilibration}
During the equilibration phase it is not necessary to keep track of all
clusters. Nevertheless there is some statistics, which is very cheap to
gather: The distribution of burnt clusters and the density of trees. The
latter is very simple, as this number changes in time only by the
number of grown trees minus the number of burnt trees. This is also a
crosscheck for the overall statistics, as the tree density is equivalent
to the probability of a site to belong to {\it any} cluster
\eref{def_rho}. 

The burning is implemented as follows: The master chooses a site from
the entire lattice and sends the corresponding slice (slave) the
coordinate and (implicitly) an identifier which uniquely identifies this
request within this update step. The slice's response consists of the
number of sites burnt (possibly $0$), the identifier referring to the
initial request and possibly up to two further, new, unique
identifiers. These identifiers refer to the two possible sub-requests to
the right and left neighbouring slice due to a spreading of the fire. If
a slice contacts another slice, it does so by sending the coordinates of
sites, which are on fire in the sending patch, together with a unique
identifier. The slice contacted sends it result to the master, again
together with the identifier and possibly two new ones, corresponding to
the possibly two contacted neighbouring slices. In this way the master
keeps track of ``open (sub)requests'', i.e. requests the master has been
told about by receiving an answer containing information about
sub-requests, which have not been matched by receiving a corresponding
answer. The structure of requests forms a tree-like structure and if
there are no open requests, the master must have received all answers of
the currently burning fire. It is very important to make it impossible
that by a delay of messages some answers are not counted, as it would
be, if the master would just count open requests, without identifying
them individually. It can easily happen that the master receives an
answers for a request, without having received the information about the
very existence of the request. It is worth to mention that in this
scheme the order of burnings is irrelevant, if the burn-time is not
measured, as it was done here.

Adding up the number of burnt sites gives the total size of the
burnt cluster. This number is finally sent to all slices. If it is
nonzero, the step is considered as successful.

After equilibration the cluster structure of pointers and roots as
described above (see section \ref{sec:tracking_clusters}), needs to be
constructed. This is done in a na\"{\i}ve manner: Keeping track of
sites, which have already been visited, every site is visited once. The
first site visited in each cluster becomes root of all sites connected
to it, which become marked as visited. The procedure corresponds to the
burning procedure described above (see section
\ref{sec:burning_procedure}).

Each slice maintains a local histogram $\PCA$, which contain all
clusters, which do not have a site on the border to another
slice. Otherwise, they are maintained at the master's histogram, as
discussed below. In this case the
(local) root site of these clusters are moved to the border. As periodic
boundary conditions apply, the only boundaries are those with other slices.

\subsubsection{Collecting statistics}
After finishing the equilibration phase another concept needs to be
applied in order to count the total cluster size distribution
$\PCA_t(s)$. At every update of the lattice each slice must keep track
of the clusters in the same way as it was described in section
\ref{sec:tracking_clusters}. Clusters, which do not contain a site at a
border to another slice are maintained locally, i.e. at each node as a
\emph{local histogram}. However, if a cluster contains a site at a
border, it might span several slices. As soon as a cluster acquires a
site at the border, it is removed from the local histogram and the
site under consideration becomes the root of the cluster. The algorithm
ensures that a cluster with at least one site on the border has its root
at the border.

During all processes (growing or burning), the size of all clusters is
updated as usual, independent from the location of the root.  If the
status of a border site changes, its new value or its change is put on a
stack together with its coordinate.  During the growing procedure the
following changes of the status are possible:
\begin{itemize}
\item \emph{New occupation:} Change in occupation information for a
      site (cell); If this is the only change, then it must have been
      already occupied (this is only possible, in an implementation
      using cells). If this is not the case, the reference information
      pointing to the root site of the given cluster, must be updated
      also, see next point.
\item \emph{Merging border clusters:} Change of the reference
      information for a site (cell); This can only happen if the site
      (cell) was (completely) unoccupied at the time of the change or
      did contain a size information, i.e. it was itself a root.
\item \emph{General merging of clusters:} Change in size information for
      a site (cell); Only an increase is possible, so that any change
      can be represented by a single number indicating the size
      difference.
\end{itemize}

For each border site changing at each slice, the corresponding
information are sent to the master. Typically the number of messages is
not very large, because the total number of sites updated during a
single growing phase is limited by $\INFL/S$. The expected number of
these messages is not given by the fraction of border sites in each slice,
because changes in all border \emph{clusters} (i.e. clusters with at
least one site in the border) affect the border \emph{sites}, as the
root of each border cluster is a border site.

However, the data regarding the updates in the border do not need to be
send from the slaves to the master, if the burning attempt following the
growing fails, i.e. if an empty site has been selected for lightning. Of
course it is much more efficient not to send any data if not
necessary. As there is only a finite number of sites in each slice, the
theoretical limit of updates of border sites is bound by this
number. However, it is sufficient to allocate a reasonable amount of
memory ($2 L$ turned out to be enough) for the stack of messages to be
sent and check its limits, similar to the stack used in the burning
procedure described in section \ref{sec:burning_procedure}. Henceforth
the sending of the update information of the border is called ``sending
the border''.

The master maintains a copy of the state of the border sites and updates
a \emph{global histogram} of border clusters. By sending the changes on
the border to the master as described above, the master can update its
copy of the configuration of the borders as well as the global
histogram. At the end of the simulation all histograms ($S$ slaves
histograms plus the global histogram maintained by the master node) are
summed to produce the total $\PCA$.

\begin{figure}[th]
\begin{center}
%% 6.5 * 0.8
\begin{pspicture}(0,0)(8.0,5.5)
%\begin{pspicture}(0,0)(11.2,7.7)
%\psset{xunit=1.4cm,yunit=1.4cm}
%\rput(0.4pt,0){\psline[linestyle=dashed](0.9,0.5)(7.5,0.5)}
%\rput(0.4pt,0){\psline[linestyle=dashed](0.9,5.0)(7.5,5.0)}

\psframe[fillstyle=solid,fillcolor=white,linestyle=dashed](0.5,0.5)(7.9,5.0)

\psframe[fillstyle=solid,fillcolor=white,linecolor=white](2.9,0.5)(3.0,5.0)
\psframe[fillstyle=solid,fillcolor=white,linecolor=white](5.4,0.5)(5.5,5.0)

%\psframe[fillstyle=solid,fillcolor=white](0.5,0.5)(0.9,5.0)
\psframe[fillstyle=vlines,hatchangle=-45,fillcolor=black](0.5,0.5)(0.9,5.0)
\rput(1.7,3.0){{\large slice 0}}
\psframe[fillstyle=vlines,hatchangle=-45,fillcolor=black](2.5,0.5)(2.9,5.0)

\psframe[fillstyle=vlines,hatchangle=-45,fillcolor=black](3.0,0.5)(3.4,5.0)
\rput(4.2,3.0){{\large slice 1}}
\psframe[fillstyle=vlines,hatchangle=-45,fillcolor=black](5.0,0.5)(5.4,5.0)

\psframe[fillstyle=vlines,hatchangle=-45,fillcolor=black](5.5,0.5)(5.9,5.0)
\rput(6.7,3.0){{\large slice 2}}
\psframe[fillstyle=vlines,hatchangle=-45,fillcolor=black](7.5,0.5)(7.9,5.0)

\psbezier{->}(0.7,1.0)(1.5,1.3)(1.5,2.0)(0.7,2.0)
\psbezier{<-}(0.7,1.0)(1.0,0.3)(1.5,0.6)(2.7,1.3)

\psbezier{->}(0.7,2.5)(2.0,1.3)(2.5,2.6)(2.7,2.7)
\psbezier{->}(2.7,2.7)(2.2,3.0)(2.2,3.2)(2.7,3.5)
\psbezier{->}(0.7,3.8)(2.0,4.3)(2.5,4.6)(2.7,3.5)

\psbezier{->}(3.2,4.8)(3.9,4.3)(3.9,4.6)(3.2,3.5)
\psbezier{->}(3.2,1.8)(3.8,2.3)(3.8,2.6)(3.2,3.5)
\psbezier{<-}(3.2,1.8)(3.8,1.3)(4.2,1.0)(5.2,0.7)

\psbezier{->}(5.2,3.5)(4.7,3.8)(4.5,4.5)(5.2,4.5)

\psbezier{->}(5.7,3.5)(5.7,3.8)(6.5,4.5)(7.7,4.5)
\psbezier{->}(7.7,4.5)(7.2,3.8)(7.2,3.5)(7.7,3.0)
\psbezier{->}(7.7,3.0)(6.7,2.0)(6.2,1.5)(5.7,1.5)

%\rput(5.7,0.7){$B$}
%\rput(5.7,4.8){$A$}
%\rput(7.7,0.7){$D$}
%\rput(7.7,4.8){$C$}

\rput(5.7,0.2){$B$}
\rput(5.7,5.3){$A$}
\rput(7.7,0.2){$D$}
\rput(7.7,5.3){$C$}

\psline{|<-}(0.2,0.5)(0.2,2.6)
\psline{->|}(0.2,3.4)(0.2,5.0)
\rput(0.2,3.0){$L$}
\end{pspicture}
\end{center}
\caption{ \flabel{border_copy} The slices, three of which are shown
here, maintain the references for all clusters within each slice
(illustrated by arrows), even for border clusters. The references
\emph{between} slices, however, are maintained by the master. The
variables $A=0$, $B$=L-1, $C=I$ and $D=I+L-1$ are the indices used for
references within each slice.}
\end{figure}
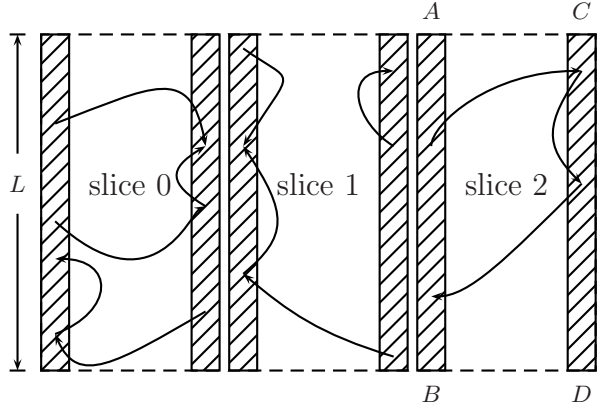

As suggested in \fref{border_copy}, the slices maintain the
pointers within each slice and these references are not changed by the
master, which only connects \emph{between} slices. If a reference at the
border changes at a slice, the master receives a message to apply the
corresponding changes (joining two clusters), if the size of a cluster
changes, the master updates the corresponding unique root etc. These
changes are indicated by the slaves and the master only realises them in
the copy of the border sites. Only if a change in occupation occurs, the
master must actually perform some non-trivial operations, because a
newly occupied site might introduce a new connection \emph{between}
borders of different slices. From the point of view of the master, only
borders belonging to two different, neighbouring slices are directly
connected and therefore to be maintained by the master, while the
connectivity of the borders \emph{within} each slice is indicated and
maintained by the corresponding slave. Apart from that, the master
maintains the slice spanning structures in exactly the same way as the
slaves, e.g. a cluster having multiple roots among the various slices
has a unique root at the master etc.

The question arises how the master best keeps track of the changes of
the borders. Ideally, a change of reference of a site at the boundary is
communicated to the master simply by sending the new pointer value
(index). By choosing a reasonable indexing scheme, this is indeed
possible. If the value of the reference is within $0$ and $L-1$, where
$L$ is the width in terms of number of sites (or cells) (see
\fref{border_copy}), the reference denotes a site in the left border
within the same slice. Similarly, if the value of a reference is within
$I$ and $I+L-1$, where $I$ denotes the first index in the last column, a
reference with such a value is bound to point to the right border of the
same slice.  If the master uses indexes of the range $[L,I-1]$ for
denoting cross-references between slices the references are therefore
unambiguous and no translation is necessary between indeces used by the
slices and indeces used by the master.

During the burning procedure the master can make use of its knowledge
about the borders. The site selected for starting the fire is most
likely a bulk size, so that the corresponding slave needs to be
contacted for the occupation information. Three outcomes are possible:
\begin{itemize}
\item The site is unoccupied. Nothing happens, all slices get signalled
      to continue with growing.
\item The site is occupied, but does not contain a border site. In this
      case the slice contacted can send back the size of the burnt
      cluster (an information it knows even without actually doing the
      burning as the size is stored in the root, which needs to be found
      anyway in order to find out whether the cluster is a border
      cluster) and the master can signal all other slices to send the
      border and to continue. After receiving the borders it can update
      the histogram.\footnote{One might be inclined to postpone the
      sending of the borders to a time, when it is really
      needed. However, after a successful burning the time $t$ is
      increased and this enters the histogram (see section
      \ref{sec:efficient_histogram_superposition}). Ignoring this change
      for a large number of steps would introduce uncontrollable
      deviations of the estimator of the histogram from its true value.}
\item The site is occupied and contains a border site. In this case the
      slice sends the reference of the border site back to the master,
      which then contacts all slices to send the most recent border
      update. It updates the border and the histogram, deletes the
      cluster which is going to burn and sends the ``burning borders'',
      i.e. a list of all border sites which will be affected by the
      burning procedure to the slices in form of a stack as described in
      section \ref{sec:burning_procedure}. The slaves use this stack as
      the initial stack of the burning procedure and delete the
      corresponding sites. No communication between the slices is
      necessary.
\end{itemize}

The global histogram contains much larger clusters than the local
histograms. In order to keep memory requirements low, even for
histograms of resolution unity, it is reasonable to introduce a threshold,
above which slaves use the global histogram to maintain $\PCA$ even for
local clusters (i.e. non border cluster). For that purpose a histogram
``appendix'' has been introduced. This is a finite stack, which stores
the size of the cluster $s$ together with the value of $t'=\pm(T-t+1)$ as
described in section \ref{sec:efficient_histogram_superposition}. During
the growing phase when such large clusters grow fastly, one would obtain
a sequence of stack entries of the form $(s, t'), (s, -t'), (s+1, t'),
(s+1, -t'), (s+2, t'), \dots$, corresponding to entering the appendix,
$(s, t')$, increasing in size by $1$, which gives $(s, -t'), (s+1, t')$
etc.  As soon as a cluster is larger then the upper cutoff each update
causes two entries, of the form $(s, -t'), (s+1, t')$, the first for the
deletion from the histogram, the second from the increase in the next
slot. These entries possibly cancel, for example the sequence above is
equivalent to the single entry $(s+2, t')$. It turned out to be highly
efficient to perform this cancellation, i.e. to check the last entry in
the appendix for being the negative entry of the one to be done.

As the maximum size of the appendix is finite, it must be emptied from
time to time. The information about the size of the appendix of each
slave is sent to the master together with the information about the
borders. If a possible overflow is detected ($2/3$ of the maximum
size in the implementation presented) the master requests all slices to
send the content of their appendices and applies it to the global
histogram. The slices then empty their appendices.

\subsubsection{The Random Number Generator}
The random number generator (RNG) acquires a crucial role when used in a
parallel environment. With $M$ the number of iterations, the expected
number of calls of the RNG is $M \INFL / \rho$ (for $M \approx 10^7,
\INFL \approx 5 \times 10^4$ this is more than $5 \times 10^{11}$), so
that an RNG as {\tt ran1} in \citep{Press:92} with a period of only
$\approx 2 \times 10^9$ is insufficient. Therefore {\tt ran2} in
\citep{Press:92} was used for all simulations, both parallel and
non-parallel, which has a period of $>2 \times 10^{18}$.  If the number of
RNG calls is small enough, one can compare results obtained by means of
{\tt ran1} and {\tt ran2}. No significant deviation was found.

In the parallel implementation, each slave requires an independent
sequence of random numbers. This is a classical problem in parallel
computing \citep{AluruPrabhuGustafson:1992,Coddington:1996}. The simplest
solution is to divide a single sequence $r_1, r_2, \dots$ into distinct
subsequences. This can be done either by a leapfrog scheme
\citep{Coddington:1996,Entacher:1999}, where each subsequence consists of
random numbers which are $S$ calls away, i.e. $S$ subsequences of the
form $r_u, r_{S+u}, r_{2S+u}, \dots$ with $u=1, 2, \dots, S$ unique at
each slave, or by splitting the sequence \citep{Coddington:1996}, so that
each subsequence consists of consecutive RNG calls, i.e. $r_{1+u X},
r_{2+u X}, r_{3+u X}$ again with $u=1, 2, \dots, S$ and offset $X$ large
enough to avoid any overlap. The latter scheme has the advantage that
the sequence consists of consecutive RNG calls and therefore has been
used in the following. The implementation of the offset $X$ at each
slave is easily realised by restoring all state variables of the RNG,
which have been produced once and for all in a single run producing all
$X S$ random numbers and saving the state variables on a regular
basis. However, such a technique is advisable only if the RNG calls do
not dominate the overall CPU time, in which case it would take almost as
long as the simulation itself to produce the random numbers required for
it.

\section{Results}

\begin{figure}[t]
\begin{center}
\includegraphics[width=0.7\linewidth]{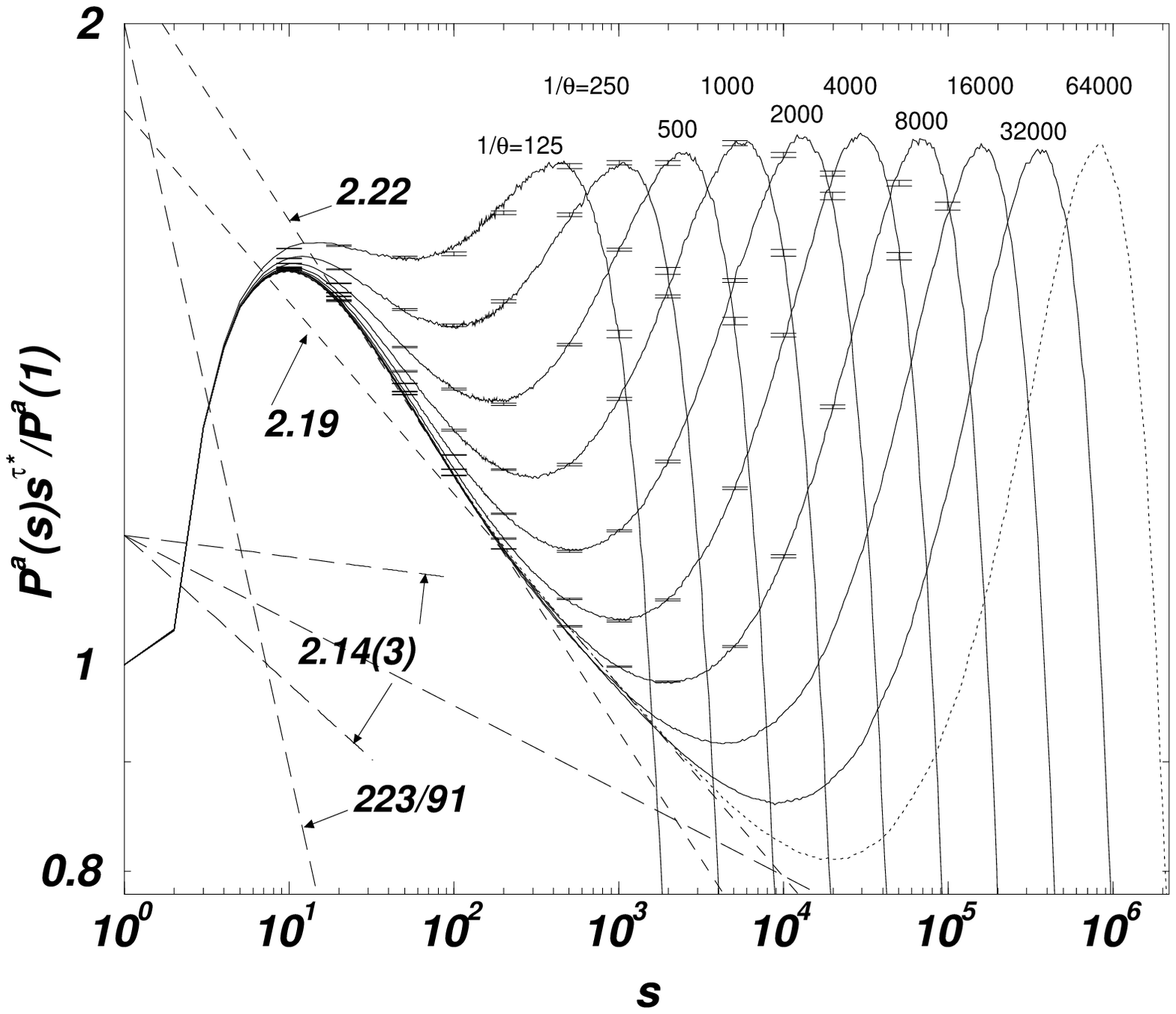}
\caption{\flabel{scaling_function} 
The rescaled and binned histogram
%$\frac{\PCA(s)}{\PCA(1)} s^{\Tast}$,  
$\PCA(s) s^{\Tast}/\PCA(1)$,
where $\Tast=2.10$ 
for $\INFL=125, 250, 500, \dots, 32000, 64000$ (as indicated) in a double logarithmic
plot. The linear size $L$ is chosen according to the bold printed
entries in Tab.~\ref{tab:absolute_results} and large enough to ensure
absence of finite size effects. The error-bars are estimated from
shorter runs. The rightmost histogram (dotted,
$\INFL=64000$) could not be cross-checked by another run, see
text. 
The dashed lines belong to different exponents, whose value is specified
as the sum of the slope in the diagram and $\Tast$, i.e. a horizontal line
would correspond to an exponent $2.1$. The shortly dashed lines
represent estimated exponents for different regions of the histogram ($2.22$
for $s$ within approx. $[20,200]$ and $2.19$ for $s$ within $[200,2000]$), the other
exponents are from literature, namely $2.14(3)$ in
\citep{ClarDrosselSchwabl:1994,ClarDrosselSchwabl:1996} and
$223/91\approx 2.45$ in \citep{Schenk:2002}. Since it was impossible to
relate these exponents to any property of the data, the exact position
of the lines associated with them was chosen arbitrarily.  }
\end{center}
\end{figure}

\begin{table}
\caption{\label{tab:absolute_results} Parameters and results for different
choices of $L$ and $\INFL$.  The average cluster size is denoted by
$\aves{s}$, for definition see \eref{ave_s}, but due to a truncation
in the histogram for some of the simulations in the range $2000\le \INFL
\le 16000$, the number presented is actually the average size of the
burnt cluster. In the stationary state it is --- apart from small
corrections --- also given by $(1-\rhobar)/(\theta \rhobar)$, see \eref{aves_averho}.
Values of $\INFL$ and $L$ printed in
bold indicate results shown in \fref{scaling_function}, the
other results are only for comparison.  All data are based on $5\times
10^6$ (successful) updates (s. Sec.~\ref{sec:clusterdistribution}) for
the transient and statistics, apart from those printed in italics which
are based on short runs ($5\times 10^6$ updates for the transient and
$1\times 10^6$ updates for statistics).
}
\newcolumntype{d}[1]{D{.}{.}{#1}}
\begin{tabular}{r|r|d{5}|d{2}|d{6}|d{2}}
$\INFL$    & $L$        & 
\multicolumn{1}{c|}{$n(1)$} &
\multicolumn{1}{c|}{$\aves{s}$} & 
\multicolumn{1}{c|}{$\rhobar$} &
\multicolumn{1}{c}{$\frac{1-\rhobar}{\theta \rhobar}$}
 \\
\hline 
{\it 125}    & {\it 1000}                                           & 0.04553     & 204.07     & 0.37973 & 204.18  \\
     125     &      1000                                            & 0.04552     & 203.81     & 0.37977 & 204.15   \\
{\it 125}    & {\it 4000}                                           & 0.04553     & 203.88     & 0.37983 & 204.10  \\
{\bf 125}    & {\bf 4000}                                           & 0.04552     & 203.77     & 0.37983 & 204.10   \\
\hline
{\it 250}    & {\it 1000}                                           & 0.04451     & 395.03     & 0.38756 & 395.06  \\
     250     &      1000                                            & 0.04452     & 394.08     & 0.38750 & 395.15   \\
{\it 250}    & {\it 4000}                                           & 0.04454     & 394.97     & 0.38766 & 394.89  \\
{\bf 250}    & {\bf 4000}                                           & 0.04454     & 395.29     & 0.38765 & 394.91   \\
\hline
{\it 500}    & {\it 1000}                                           & 0.04380     & 764.73     & 0.39316 & 771.75  \\
     500     &      1000                                            & 0.04380     & 764.81     & 0.39315 & 771.77 \\
{\it 500}    & {\it 4000}                                           & 0.04382     & 771.12     & 0.39343 & 770.88  \\
{\bf 500}    & {\bf 4000}                                           & 0.04382     & 771.90     & 0.39343 & 770.87  \\
\hline
{\it 1000}   & {\it 1000}                                           & 0.04328     & 1495.36   & 0.39716 & 1517.91  \\
     1000    &      1000                                            & 0.04328     & 1490.05   & 0.39714 & 1518.00    \\
{\it 1000}   & {\it 4000}                                           & 0.04331     & 1510.85   & 0.39761 & 1515.00   \\
{\bf 1000}   & {\bf 4000}                                           & 0.04331     & 1513.13   & 0.39764 & 1514.81    \\
{\it 1000}   & {\it 8000}                                           & 0.04332     & 1510.10   & 0.39763 & 1514.91   \\
\hline
{\it 2000}   & {\it 4000}                                           & 0.04296     & 2976.34    & 0.40053 & 2993.35  \\
{\bf 2000}   & {\bf 4000}                                           & 0.04297     & 2990.50    & 0.40054 & 2993.15   \\
{\it 2000}   & {\it 8000}                                           & 0.04297     & 2995.67    & 0.40060 & 2992.56  \\
\hline
{\it 4000}   & {\it 4000}                                           & 0.04273     & 5929.24     & 0.40258 & 5935.91 \\
     4000    &      4000                                            & 0.04273     & 5930.97     & 0.40249 & 5938.03  \\
{\it 4000}   & {\it 8000}                                           & 0.04274     & 5931.32     & 0.40261 & 5935.15 \\
{\bf 4000}   & {\bf 8000}                                           & 0.04273     & 5935.36     & 0.40256 & 5936.47  \\
\hline
{\it 8000}   & {\it 4000}                                           & 0.04255     & 11786.97    & 0.40405 & 11799.72 \\
     8000    &      4000                                            & 0.04255     & 11788.90    & 0.40406 & 11799.07  \\
{\it 8000}   & {\it 8000}                                           & 0.04257     & 11801.31    & 0.40412 & 11795.98 \\
{\bf 8000}   & {\bf 8000}                                           & 0.04257     & 11792.82    & 0.40413 & 11795.38 \\
\hline
{\it 16000}  & {\it 4000}                                           & 0.04244     & 23430.01    & 0.40525 & 23481.82 \\
{\it 16000}  & {\it 8000}                                           & 0.04243     & 23466.93    & 0.40540 & 23467.22 \\
     16000   &       8000                                           & 0.04243      &  23446.10   & 0.40542 & 23465.64   \\
{\bf 16000}  & {\bf 16000}                                          &  0.04245      & 23449.31   & 0.40541 & 23466.57   \\
\hline
     32000   &      16000                                           & 0.04232     &  46443.83    & 0.40660 & 46701.82 \\
{\bf 32000}  & {\bf 32000}                                          &  0.04233    &   46731.44   & 0.40662 & 46698.51 \\
\hline
{\bf 64000}  & {\bf 32000}                                          &
 0.04220     &   91148.64  & 0.40777 & 92952.40 \\
\end{tabular}
\end{table}

The sections above were only concerned with the technical issues of the
model and its implementation. Some of the actual results from the
simulation carried out using the new algorithm have been published
already \citep{JensenPruessner:2002b}. This article was focused on
$\dns$. The main outcome was that the standard scaling assumption
\eref{def_tau} is not supported by numerics, so the main conclusion was
that the model \emph{is not scale invariant}.

In the following these results are shortly restated and
discussed. Other observables are connected with this observation to see,
whether it is only $\dns$ which lacks scale-invariance. All results
presented are based on the same simulations, the parameters of which are
given in Tab.~\ref{tab:absolute_results}.

\subsection{Cluster size distribution}\label{sec:clusterdist}
Before the actual findings are discussed, it is important to consider
how to avoid finite size effects, which otherwise might damage the results. 
Usually, finite size effects are avoided by keeping the correlation
length $\xi$ small compared to the system size. However, it requires a
significant amount of CPU-time to actually determine the correlation
length. Moreover, \emph{a priori} it would not be clear, which ratio
$\xi/L$ to choose in order to avoid finite size effects.

The simplest way to determine whether finite site effects are present
is to compare estimates of observables for two systems with the same
parameters but different sizes \citep{SchLoiPru:2001}. If finite size
effects are not present, the differences between the estimators of those
two systems must be within the error bar of the quantity under
consideration. This approach has the drawback that each set of
parameters must be simulated at least twice, but it gives full control
over finite size effects. Apart from $\INFL=64000$, which is specially
marked in most of the plots, this approach has been applied throughout the
results presented. The method was discussed in greater detail in
\citep{JensenPruessner:2002b}.

\begin{figure}[t]
\begin{center}
\includegraphics[width=0.7\linewidth]{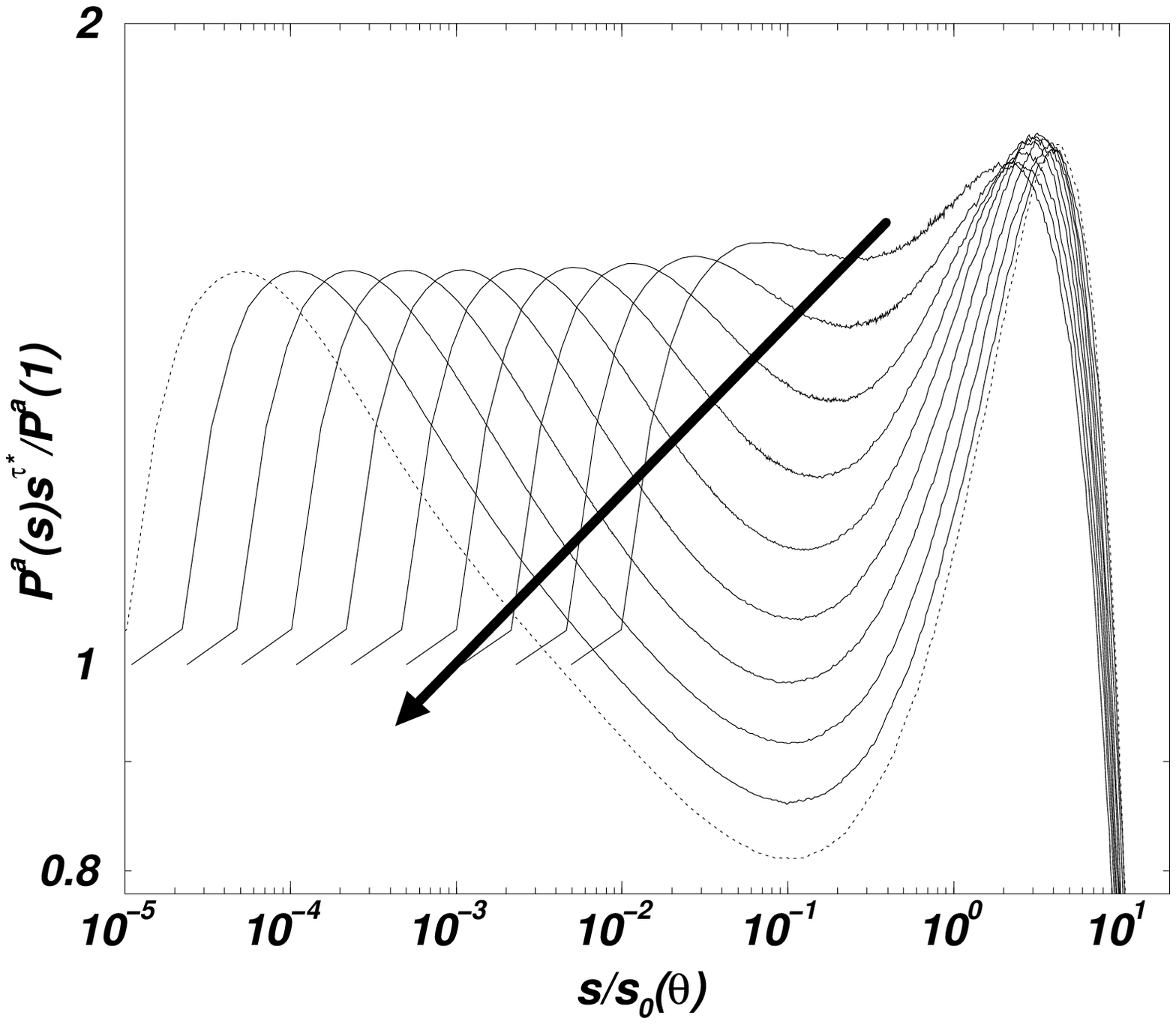}
\caption{\flabel{data_collapse} 
Attempt to collapse the data shown in \fref{scaling_function}
 using $\Tast=2.10$, $\Scutoff(\theta) = \theta^{-\lambda^\ast}$ and
 $\lambda^\ast=1.11$ as derived from
 \Eref{scaling_relation_ltau}. As expected the data do not
 collapse. 
The big arrow points in the direction of increasing $\INFL$.}
\end{center}
\end{figure}

\Fref{scaling_function} shows a central result of
\citep{JensenPruessner:2002b}. This figures contains the reduced (and
binned) data in the form
\begin{equation} \elabel{pnorm}
\frac{\PCA(s)}{\PCA(1)} \quad ,
\end{equation}
which has the convenient property to be unity for $s=1$. The
normalisation $\PCA(1)$, which converges anyway to a finite value as
$\INFL \to \infty$ (see Tab.~\ref{tab:absolute_results}), does not
affect any of the results, especially not the (attempted) data
collapses.

The crucial problem shown in \fref{scaling_function} is the intermediate
minimum that develops as $\INFL$ is increased. It renders the data
collapse as described by \eref{def_tau} impossible (for more details see
\citep{JensenPruessner:2002b}). \Fref{data_collapse} shows the same data
again, now in an attempt to form a data collapse, using
$\Scutoff(\theta) = \theta^{-\lambda^\ast}$ with $\lambda^\ast=1.11$
from \Eref{scaling_relation_ltau} and $\Tast=2.10$ (for comparison see
Tab.~\ref{tab:expos_literature}). As expected the collapse fails.

In less technical terms, it was shown in \citep{JensenPruessner:2002b}
that there is no choice of $\tau$, which allows a data collapse for
$\PCA(s; \theta)$. It seems that the distribution is the same for two
different values of $\theta$ up to a certain cluster size, which
increases seemingly unbound with $\INFL$, i.e. for two very large values
of $\INFL$ the two distributions collapse without any rescaling. Beyond
this cluster size the distributions deviate, the one with the larger
$\INFL$ forms a deeper dip and ascends afterwards to a maximum, which
can, by rescaling, be arranged to be the same for all $\INFL$. The ever
growing dip prohibits a reasonable definition of a lower cutoff and
makes a data collapse impossible. Equally one could arrange the dips to
be at the same height and the maximum to increase in $\INFL$.

The key problem of the DS-FFM is that more than one length scale is
visible apparently for any system size $L$. The statistics of $\dns$ is
not even asymptotically dominated by a single length scale. For any
system size, a $\dns$ only given for all $s$ larger than any lower
cutoff, allows the identification of $\theta$ by the shape of $\dns$
alone. 

This indicates that simple scaling \eref{def_tau} does not apply and
the exponent $\tau$ is undefined. Keeping this in mind, it is very
instructive to look for other properties as well and investigate their
scaling.

\subsubsection{Finite size scaling} \label{sec:finite_size_scaling}
The failure of the DS-FFM to obey proper finite size scaling has been
observed in \citep{SchenkETAL:2000} already. In the following some finite
size scaling principles have been applied in a straight forward manner
and subsequently ruled out.

As known from percolation \citep{StaufferAharonyENG:1994}, the
generalised form of the scaling behaviour of $\Scutoff$ is 
\begin{equation} \elabel{scutoff_generalised}
 \Scutoff(\theta, L) = \theta^{-\lambda} m\left(\theta L^\sigma \right)
\end{equation}
where $m(x)$ is a crossover function describing the dependence of
$\Scutoff$ on the two parameters $\theta$ and $L$. For sufficiently
large argument $x$, the crossover function is expected to approach a
constant, such that \eref{def_scutoff} is recovered. For small
arguments, however, the dependence of the cutoff is expected to be
strongly dominated by $L$, just like in equilibrium critical phenomena,
where $L$ takes over the r\^ole of $\xi$ for sufficiently small systems.
Thus, for small arguments $m(x) \propto x^\lambda$, so that for
sufficiently small $L$, $\Scutoff$ becomes independent of $\theta$.

Generic models of SOC do not have any tuning parameters other than the 
system size, so that the cutoff $\Scutoff$ is only a function of $L$. In this
sense, finite size scaling is the only scaling behaviour in SOC, and a
failure of the model to comply to finite size scaling is identical to
the failure to comply to simple scaling altogether. Therefore, one might be
surprised to see a simple scaling analysis \emph{and} a finite size
scaling analysis in an article on an SOC model. However, the Forest Fire
Model is different in this respect, as it has the additional parameter
$\theta$, which is, supposedly, finite only because of the finiteness of
the system size. In the thermodynamic limit it supposedly disappears as
a free parameter.

As seen above (see \Fref{data_collapse}), the $\theta$-dependence of $\dnst$ can not be captured by
$\Scutoff$ in the scaling function alone. However, the scaling form
\eref{def_tau} would remain valid in some sense, if in the finite size
scaling regime the $L$ dependence of $\dnst$ enters $\Scutoff$
only. Therefore the original form \eref{def_tau} is generalised to
\begin{equation} \elabel{scaling_generalised}
 \dn(s; \theta, L) = s^{-\tau} \GC(s/\Scutoff(\theta, L))
\end{equation}
ignoring that it has been shown above already that it does not hold in
the limit where $\dn(s; \theta, L)$ becomes independent of $L$. In this
section the dependence of $\dn(s; \theta, L)$ on $L$ is investigated, in
the limit of large $\INFL$ and small $L$. A similar study has been
performed by Schenk \etal \citep{SchenkETAL:2000}, however on much
smaller scales and using $\PCB$.

If the form \eref{scaling_generalised} holds, it should be possible
to collapse $\dn(s; \theta, L)$ for different $L$ by choosing the correct
$\tau$ and $\Scutoff$, just like for the cluster size distribution
of standard percolation. This turns out
not to be the case, as can be seen in
\fref{finite_size_scaling_theta}: The \emph{smaller} $L$ is, the \emph{stronger}
the changes of shape of $\dns$ for any $\theta$
tested. Consequently, \eref{scaling_generalised} does not hold, and
as $\Scutoff$ is only \emph{defined} via its r\^ole as cutoff in
\eref{scaling_generalised}, $\Scutoff$ is undefined and
\eref{scutoff_generalised} remains meaningless.

One might argue that the average density of trees, $\rhobar$ (see
\eref{def_rho}), is the relevant parameter of $\dns$, so that $\dns$
has the same shape for different, sufficiently small $L$ and constant
$\rhobar$. However, as shown in \fref{density_theta}, for any value of
$\theta$, there is a value of $L$, such that $\rhobar$ varies considerably
with decreasing $L$. Especially, there seems to be a maximum tree
density for every system size, so that for large values of $\rhobar$, there
is a smallest system size $L$, below which this density cannot be
reached. This maximum increases monotonically with system size, so that
the maximum for every finite system size is smaller than the expected
average tree density in the thermodynamic limit, which is according to
Tab.~\ref{tab:absolute_results} larger than $0.40777$ and was recently
conjectured to be as large as $0.5927\dots$ \citep{Grassberger:2002},
namely the critical density of site percolation \citep{NewmanZiff:2000}. 
Accepting this limitation, \fref{finite_size_scaling_rho}
shows an example for three $\dns$ with roughly the same $\rhobar$ and
different $L$ and $\theta$. Most surprisingly two of the histograms
collapse already without rescaling, while the third ($L=500$) reveals
the same problems as visible in \Fref{scaling_function}. Hence, finite
size scaling does also not work for fixed $\rhobar$.

That large densities of trees cannot be reached by small system sizes
is related to the specific way the histograms are generated and the
density measured: Is it before or after each (successful) burning? For
sufficiently large systems, it becomes irrelevant when to do it, because
two histograms, one measured before, the other one right after the
burning, differ only by one cluster. Also
the question, whether to average only over successful burnings is
irrelevant, because the difference between a histogram before and after
the burning is only one cluster. 

\begin{figure}[t]
\begin{center}
\includegraphics[width=0.7\linewidth]{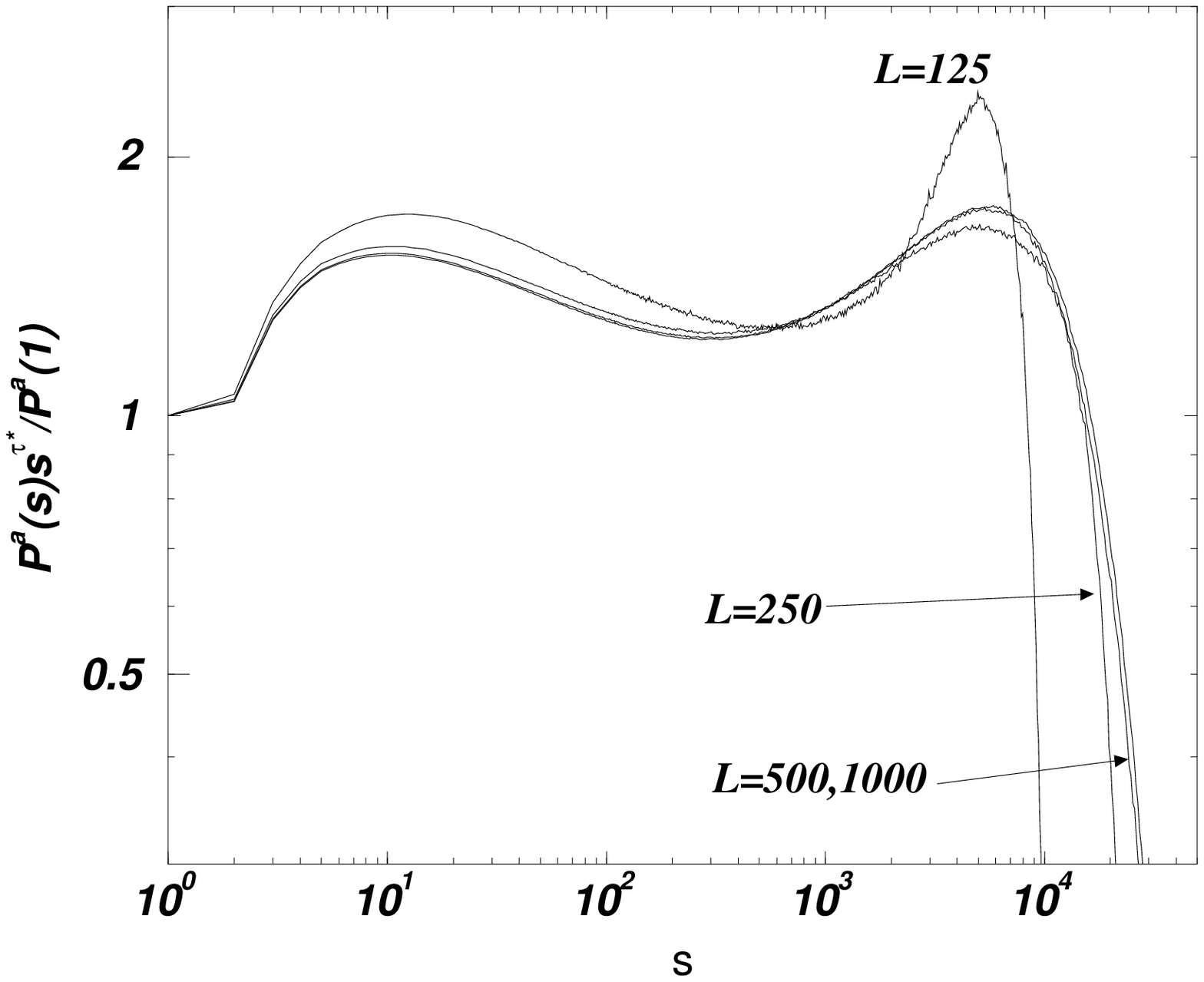}
\caption{\flabel{finite_size_scaling_theta}
Plot of the rescaled PDF $\PCA(s; \theta, L) s^\Tast/\PCA(1; \theta, L)$ for fixed $\INFL=1000$ and
different system sizes, $L=125,250,500,1000$. The different shapes make
it impossible to collapse the data, as would be expected from a finite
size scaling ansatz \eref{scaling_generalised} and
\eref{scutoff_generalised}.
}
\end{center}
\end{figure}

Clearly, for small systems, the difference between the histogram before
and after the burning, is just the one enormous cluster of size
$\OC(\INFL)$. \fref{hist_before_after} shows the difference. Even
though in principle every density is reachable for every system
size if the histogram is measured before burning, the newly defined histograms do not have a considerably different
shape, so that a collapse remains impossible. For example, the problems
shown in \fref{finite_size_scaling_theta} become even more
pronounced, if the histogram is taken before burning. 

Surprisingly and actually in contradiction to what has been said in
\Eref{converge_pcb}, there is a discrepancy between the cluster
size distribution of burnt clusters, $\PCB$, and the overall cluster
size distribution $\PCA$, even if the latter is measured \emph{before}
the burning takes place. This sounds paradoxical, because the random
picking of a cluster to be burnt is just a sampling of $\PCA$. This cannot
be caused by the correlation between those samples, due to the fact
that $n_{t+1}(s)$ is actually a function of the cluster chosen at $t$ --- a
correlation like this would be equally picked up by $\PCA$. The reason
for this discrepancy is the fact that a site picked randomly as the
starting point of the next fire is necessarily occupied. Therefore
$n_t(s)$ with a low occupation density enter $\PCB$ over-weightedly. As
low density states contain much more small clusters then large ones,
$\PCB$ overestimates the probability of small clusters. 
A sample of $\PCB$ at a low density is indistinguishable from a
sample at high denisty, while a sample for $\PCA$ trivially
contains the information about the density.
To illustrate that, one might imagine a sequence of configurations that
consists of one state, with exactly one cluster of size
$1$, and a second state, with exactly one cluster of size
$L^2$. 
The two configurations appear with a frequency such that a
cluster of size $1$ is burnt down as often as a cluster of
size $L^2$.
The resulting $\PCA$ reports that a randomly chosen site belongs
to a cluster of size $L^2$ with probability $\half$ and to a cluster of
size $1$ with probability $1/(2 L^2)$, while $\PCB$ incorrectly
reports the same probability for both cluster sizes. The problem can
actually already be spotted in \eref{converge_pcb}, which contains a
$\rho$ on the RHS, which should rather be $\rho(t)$. The problem
disappears in the limit where $\rho(t)$ hardly changes in time, i.e. in
the limit of $\INFL \ll L^2$.

It is also clear, why \eref{aves_averho} breaks down for small
systems and large $\INFL$: The average size of the burnt cluster tends
to $L^2$, while the density tends to $0$. Apparently
\eref{aves_averho} must be incorrect for $\rho<(L^2 \theta +1)^{-1}$.

\begin{figure}[t]
\begin{center}
\includegraphics[width=0.7\linewidth]{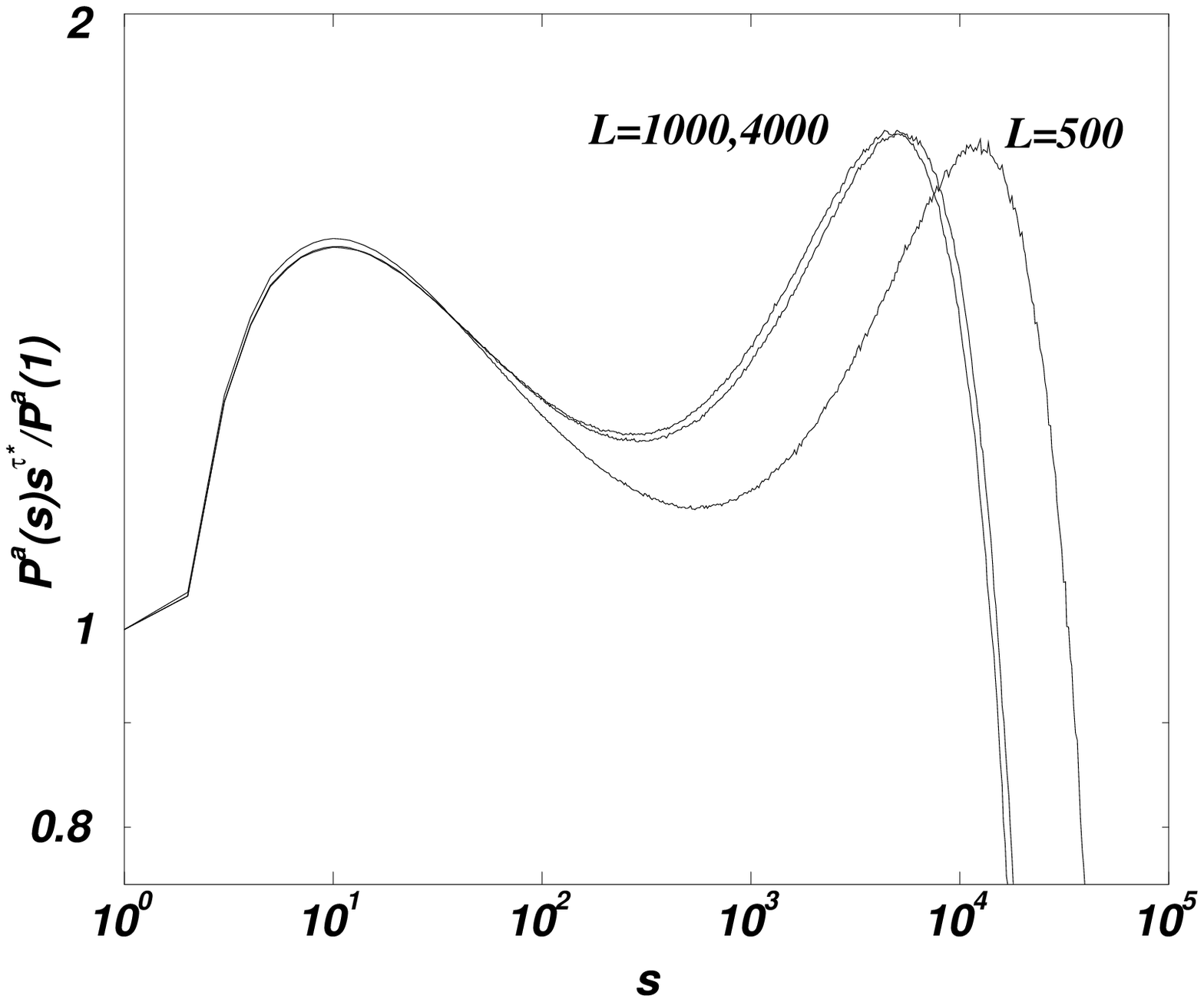}
\caption{\flabel{finite_size_scaling_rho}
Plot of the rescaled PDF $\PCA(s; \theta, L) s^\Tast/\PCA(1; \theta, L)$ for fixed
$\rhobar\approx 0.397$:
 $L=500$ with $1/\theta=2000$ ($\rhobar=0.396827$),
 $L=1000$ with $1/\theta=940$ ($\rhobar=0.396825$) and
 $L=4000$ with $1/\theta=870$ ($\rhobar=0.396883$). Again,
 a data collapse is impossible. 
}
\end{center}
\end{figure}

\begin{figure}[t]
\begin{center}
\includegraphics[width=0.7\linewidth]{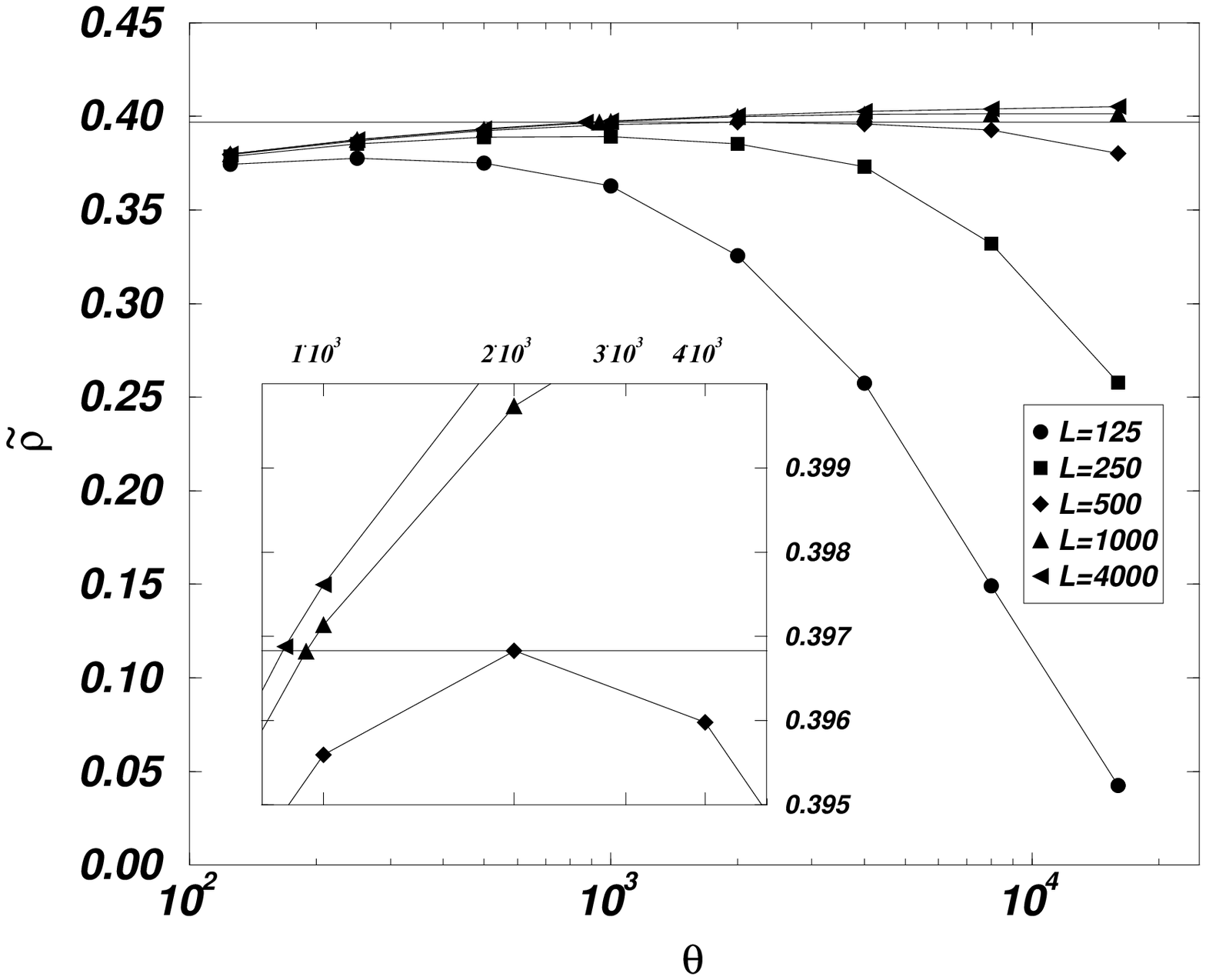}
\caption{\flabel{density_theta}
The average density of trees, $\rhobar$, as a function of $\theta$ and for
various $L$. For sufficiently small systems, the maximum in
$\rhobar$ is much smaller than the expected density at the ``critical
point'', which is larger than $0.40777$ found as in
Tab.~\ref{tab:absolute_results}. The straight line marks
 $\rho=0.396827$, the density chosen in
 \fref{finite_size_scaling_rho}. The inset is a magnification of the
 crossing of the straight line with the simulation data, and shows all
 three values of $\theta, L$ used in \Fref{finite_size_scaling_rho}.
}
\end{center}
\end{figure}

\begin{figure}[t]
\begin{center}
\includegraphics[width=0.7\linewidth]{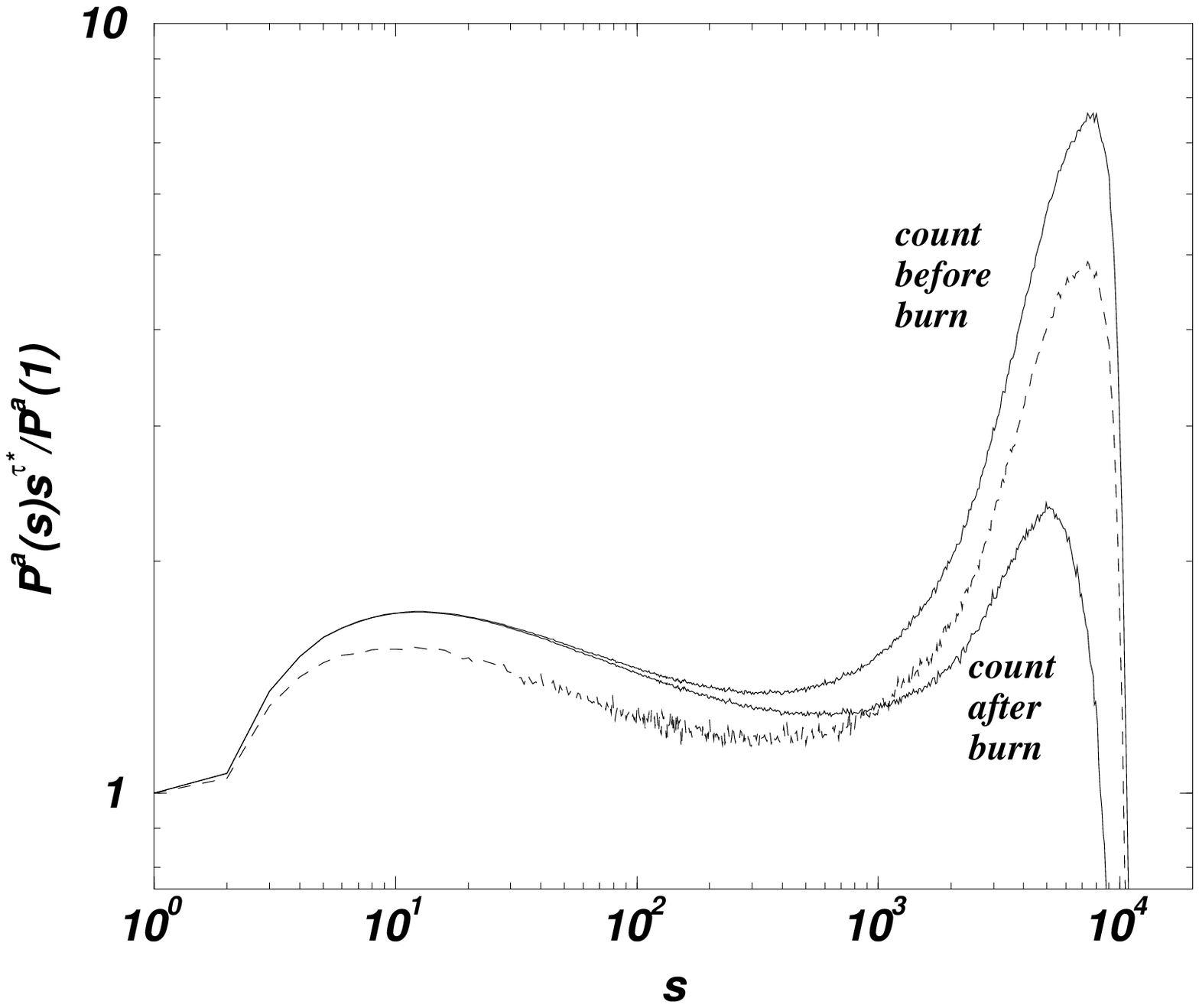}
\caption{\flabel{hist_before_after}
Comparison between the rescaled and binned histograms measured before
and after the burning for small $L=125$ and large
$\INFL=1000$. As expected, only the statistics for large $s$ is
 affected. The dashed line shows the data for $\PCB(s)$.
}
\end{center}
\end{figure}

\subsubsection{Scaling of the moments of $\PCA$}
According to \eref{def_tau}, \eref{def_scutoff} and
\eref{converge_pca} the $n$th moment of $\PCA$ should scale like
(this analysis has apparently been introduced to SOC by De Menech \etal
\citep{Pastor-SatorrasVespignani:2000,DeMenechStellaTebaldi:1998,TebaldiDeMenechStella:1999})
\begin{equation} \elabel{moment_scaling}
 \aves{s^n} = \frac{\sum_s s^n s \dnst}{\sum_{s} s \dnst} = q_n \theta^{-\lambda (2+n-\tau)} + \text{ corrections } \quad ,
\end{equation}
where $q_n$ is a non-universal amplitude (see section
\ref{sec:universal_amplitude_ratios}) and 
$\lambda$ is also known as the gap exponent \citep{Pfeuty:77}.  The
corrections are due to the lower cutoff and the asymptotic character of
the scaling, which is expected only for ``sufficiently small $\theta$''
\citep{Wegner:72}. In turn, one can infer a scaling form like
\eref{def_tau} if the moments scale in the form of
\eref{moment_scaling}.

\begin{figure}[t]
\begin{center}
\includegraphics[width=0.7\linewidth]{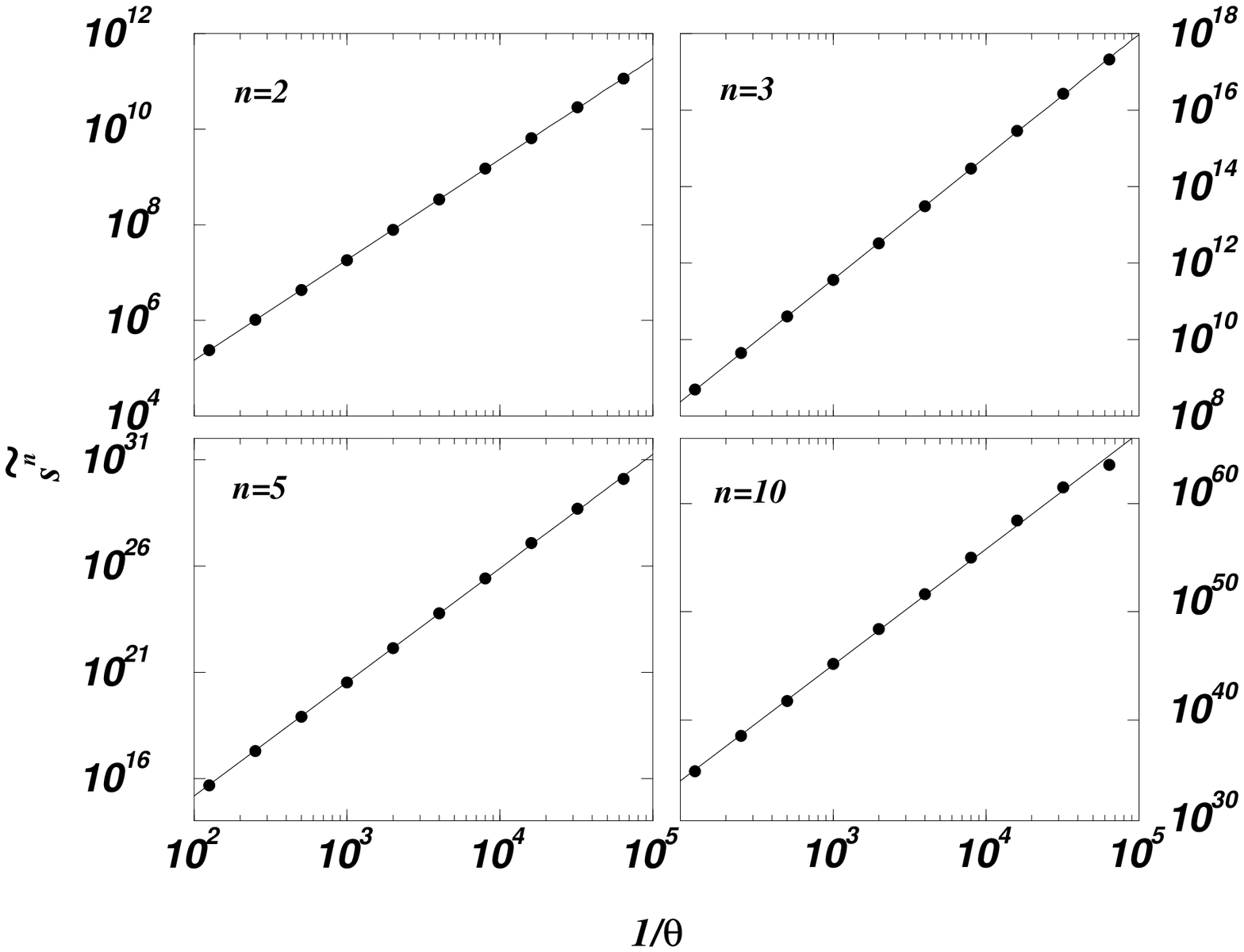}
\caption{\flabel{moment_scaling}
Scaling of the $n$th moments of $\PCA$ in double logarithmic plots. The
 straight lines show the results of a fit as $\exp(a'_n) \theta^{-\sigma_n}$, see \eref{moment_scaling}.
}
\end{center}
\end{figure}

\begin{figure}[t]
\begin{center}
\includegraphics[width=0.7\linewidth]{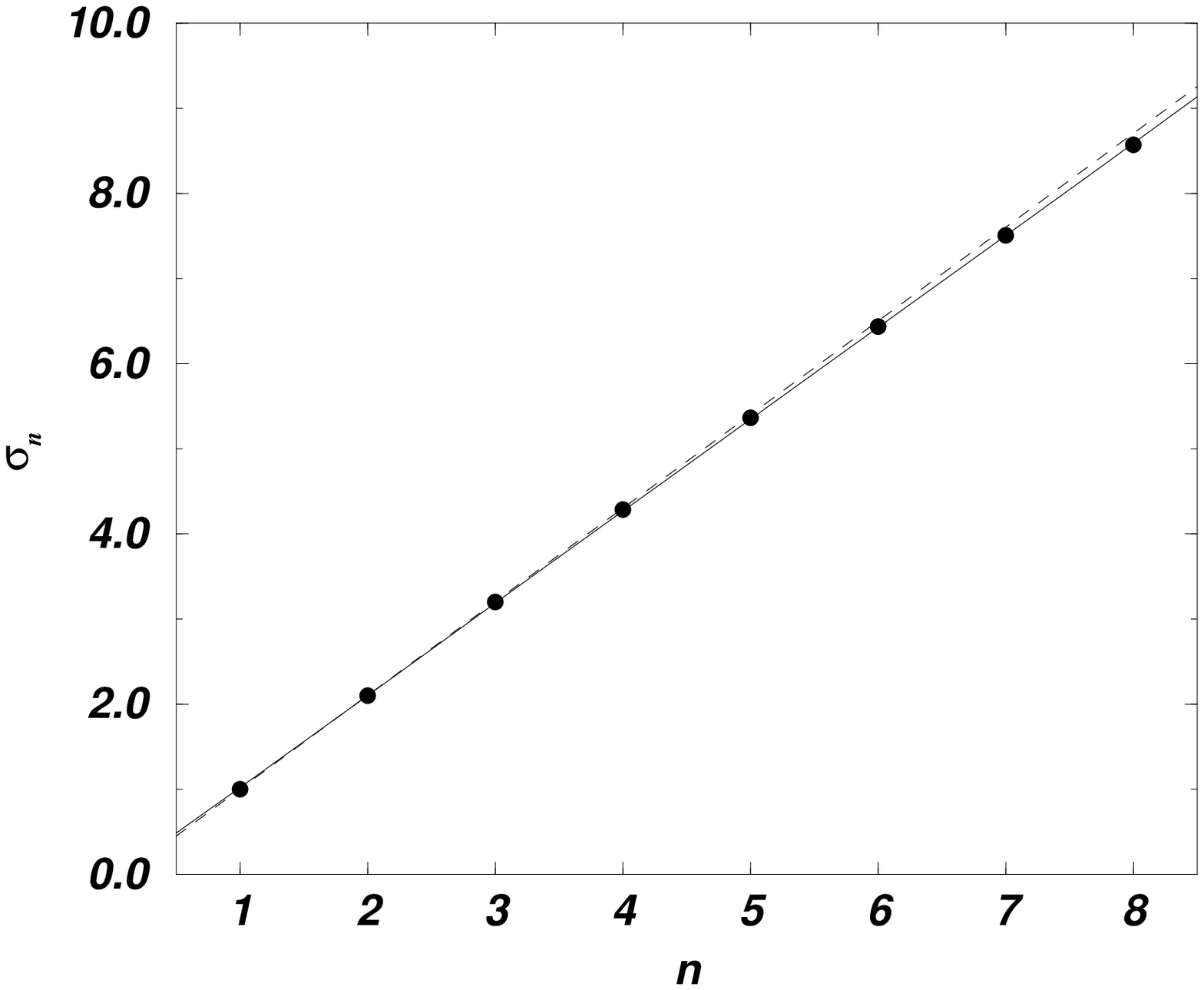}
\caption{\flabel{moment_expos}
Exponents $\sigma_n$ of the scaling of $\aves{s^n}$ in $\theta$
 vs. $n$. The slope of this curve gives $\lambda$ and $\tau$ can be derived
 from the offset. The straight, full line shows the results $\lambda=1.0808\dots$
 and $\tau=2.0506\dots$, the dashed line shows $\lambda=1.0998\dots$ and
 $\tau=2.0864\dots$ from a fit excluding $\INFL=64000$.
}
\end{center}
\end{figure}

Contrary to what is observed in an attempt of a data collapse, it turns
out that the moments follow beautifully this scaling
behaviour. \fref{moment_scaling} shows the scaling of the moments
for $n=2,3,5,10$. By simply fitting the double logarithmic data to a
straight line, i.e.
\begin{equation}
 \log(\aves{s^n}) = a'_n - \sigma_n \log( \theta)
\end{equation}
one can derive an estimate of the exponents
$\sigma_n$ and in turn compare them to the expected linear
behaviour:
\begin{equation} \elabel{expo_scaling}
 \sigma_n = \lambda (2+n-\tau) \quad .
\end{equation}
The resulting estimates, using $n=2,\dots,8$ and $\sigma_1=1$ from
\eref{ave_s}, are $\lambda=1.0808\dots$ and $\tau=2.0506\dots$,
where no statistical error is given because the systematic error, due to
neglecting of the lower cutoff as well as the corrections
\eref{moment_scaling}, is expected to be much more important. By
using the assumption $\sigma_1=1$, this result is consistent with
\eref{scaling_relation_ltau}. The results are shown in
\fref{moment_expos}.

The exponent found for $\tau$ is remarkably close to the accepted value
of standard 2D percolation, $187/91=2.054945\dots$. However, if one
leaves out the results for $\INFL=64000$, which seem to be a bit off the
lines shown in \fref{moment_scaling}, one finds a slightly larger
value for the exponent, namely $\tau=2.0864$ and
$\lambda=1.0998\dots$. This is much closer to the $\Tast=2.10$ used above. For comparison
to values found in the literature, see Tab.~\ref{tab:expos_literature}. 

It is very remarkable that the resulting estimates for the exponents are
so impressingly consistent, even though in section \ref{sec:clusterdist}
it turned out, that the scaling assumption \eref{def_tau} does not
actually hold; one would much rather expect a failure of the moments to
comply with \eref{moment_scaling}, or a failure of the exponents to
comply with \eref{expo_scaling}.  Apparently the moments are hiding
the breakdown of simple scaling. Therefore it is interesting to analyse
the behaviour of the presumably universal amplitude ratios, which are
solely a property of the (presumed) scaling function.

Another explanation for the moments being well behaved is the
following: According to \citep{JensenPruessner:2002b} one might expect the
moments to behave like
\begin{equation}
 \int_1^{\theta^{-\xmin}} ds f(s) s^n + \int_{\theta^{-\xmin}}^\infty ds s^{n-\tau} \GC(s/\theta^{-\xmax})
\end{equation}
where the first integral describes the behaviour up to the minimum, which
scales like $\theta^{-xmin}$ ($\xmin\approx0.95$) and the second
integral the behaviour from the minimum on. Because \Fref{data_collapse}
indicates already that the scaling function $\GC$ does not collapse
using a scale $\theta^{-\xmax}$ this scaling does not work and can therefore be
only an approximation. While the first integral is bound by
$\OC(\theta^{-\xmin(n+1)})$ the second integral gives $\OC(\theta^{-\xmax(1+n-\tau)})$
asymptotically, which dominates the moments for
$\xmin(n+1)<\xmax(1+n-\tau)$, which leads to $n>9.08$ using
$\xmax\approx1.2$ and $\tau\approx2.1$. \Fref{moment_scaling} shows
clearly a deviation from the straight line behaviour for $\INFL=64000$
and $n=10$ and even for $n=5$. It remains unclear whether this is due to
the effect discussed or simply a finite size problem. According to the
findings presented in section \ref{sec:universal_amplitude_ratios} the
latter might well be the case.

It is worthwhile pointing out, that the analysis in this section arrives
at estimates for the critical exponents very close to those obtained by
Pastor-Satorras and Vespignani \citep{Pastor-SatorrasVespignani:2000},
who, however, allow for the corrections in \eref{moment_scaling} which
were omitted above.

\begin{table}
\caption{\label{tab:expos_literature} Exponents of the Forest Fire Model
found in the literature. The first column indicates the source, the
second column the method. $P(s)$ denotes a direct analysis of $\dnst$,
which sometimes may have been just an estimate of the slope of $\dnst$
rather than a data collapse. For details the original sources should be consulted. The
entry ``moments'' refers to an analysis of the moments of $P(s)$, the
entry ``theoretical'' to theoretical considerations regarding the
relation of the Forest Fire Model to percolation.}
\begin{ruledtabular}
\newcolumntype{d}[1]{D{.}{.}{#1}}
\begin{tabular}{l|c|d{6}|d{7}}
reference & 
method &
\multicolumn{1}{c|}{$\tau$} &
\multicolumn{1}{c}{$\lambda$} \\
\hline 
Christensen \etal          1993 \citep{Christensen:1993}                & $P(s)$                 & 2.16(5)    & - \\
Henley                     1993 \citep{Henley:1993}                     & $P(s)$                 & 2.150(5)   & 1.167(15) \\
Grassberger                1993 \citep{Grassberger:1993}                & $P(s)$                 & 2.15(2)    & 1.08(2)   \\
Clar \etal                 1994 \citep{ClarDrosselSchwabl:1994}         & $P(s)$                 & 2.14(3)    & 1.15(3)   \\
Honecker and Peschel       1997 \citep{HoneckerPeschel:1997}            & $P(s)$                 & 2.159(6)   & 1.17(2)   \\
Pastor-Satorras Vespignani 2000 \citep{Pastor-SatorrasVespignani:2000}  & moments                & 2.08(1)    & 1.09(1)   \\
Schenk \etal               2002 \citep{Schenk:2002}                     & theoretical and $P(s)$ & 2.45\dots & 1.1       \\
Grassberger                2002 \citep{Grassberger:2002}                & $P(s)$                 & 2.11       & 1.08      \\
\end{tabular}
\end{ruledtabular}
\end{table}

\subsubsection{Universal amplitude ratios} \label{sec:universal_amplitude_ratios}
In general simple scaling involves two additional non-universal
parameters $a$ and $b$,
\begin{equation} \elabel{general_scaling}
 \dnst=a s^{-\tau} \GC\left(\frac{s}{b \Scutoff}\right) \quad .
\end{equation}
For $1<\tau<2$ the lower cutoff becomes asymptotically
irrelevant compared to the upper cutoff for all moments $n\ge 1$ ---
indeed the effective $\tau$ of $s \dnst$ fulfils this condition as
$2<\tau<3$ \citep{ClarDrosselSchwabl:1996}. Neglecting the lower cutoff then gives
for the $n$th moment of $s \dnst$
\begin{equation}
 \aves{s^n} = a (b \Scutoff)^{1+n-\tau} g_n
\end{equation}
with
\begin{equation}
 g_n \equiv \int_0^\infty dx x^{1+n-\tau} \GC(x)
\end{equation}
In order to construct universal amplitude ratios, one needs to get rid
of all exponents and parameters. This can be achieved by 
considering
\begin{equation} \elabel{uni_tmp1}
\frac{\aves{s^n}}{(\aves{s^2})^{n/2}} =  \left( a \left(b \theta^{-\lambda} \right)^{(1-\tau)}\right)^{(1-n/2)}
    \frac{g_n}{g_2^{n/2}} 
\end{equation}
and (\Eref{uni_tmp1} with $n=1$)
\begin{equation} \elabel{uni_tmp2}
\frac{\aves{s}}{\sqrt{\aves{s^2}}} = \left( a \left(b \theta^{-\lambda} \right)^{(1-\tau)}\right)^{1/2} 
    \frac{g_1}{g_2^{1/2}} \ .
\end{equation}
If one now multiples \eref{uni_tmp1} with the $n-2$th power of
\eref{uni_tmp2}, everything cancels apart from the $g_n$:
\begin{equation}
\frac{\aves{s^n}}{(\aves{s^2})^{n/2}} \frac{(\aves{s})^{(n-2)}}{(\aves{s^2})^{(n-2)/2}} =
\frac{g_n g_1^{n-2}}{g_2^{n-1}}
\end{equation}
It is worth noting that for a trivial case, where $\aves{s^n} \propto
\aves{s}^n$, the effective exponent $\tau$ is necessarily unity and
\eref{uni_tmp1} as well as \eref{uni_tmp2} are already independent
of $\theta$.

A further simplification is to impose $g_1=1$ and $g_2=1$, which fixes
the two free parameters $a$ and $b$ in \eref{general_scaling}, so that
\begin{equation} \elabel{def_g_n}
g_n = \frac{\aves{s^n}\  (\aves{s})^{(n-2)}}{\Big(\aves{s^2}\Big)^{(n-1)}}
\end{equation}
for $n\ge1$. In \fref{uni_amp} this quantity is shown for
$n=3,4,5,6$. Now, for $\INFL=64000$ a deviation is clearly visible ---
in turn that means that $\INFL=64000$ requires at least systems of the
size $L=64000$, which might explain the large value of $\rhobar$
obtained in \citep{Grassberger:2002}. Apart from that, this analysis
agrees with the result found in Sec.~\ref{sec:clusterdist}: The
supposedly universal amplitude ratios keep changing with $\theta$ and an
asymptote cannot be estimated, i.e. the scaling \eref{def_tau} is
broken. 

\begin{figure}[t]
\begin{center}
\includegraphics[width=0.7\linewidth]{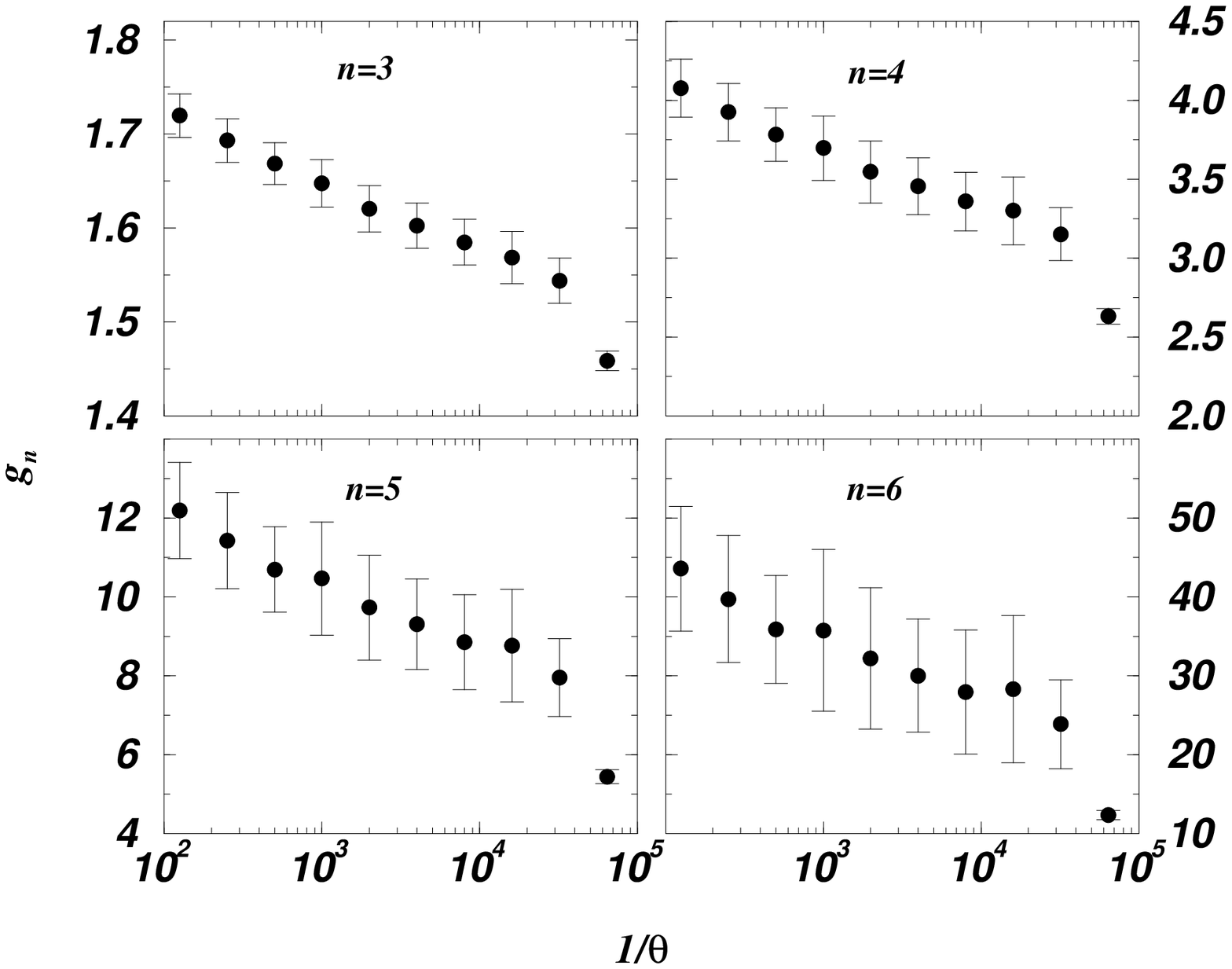}
\caption{\flabel{uni_amp}
The supposedly universal amplitude ratio $g_n$ \eref{def_g_n} for
 $n=3,4,5,6$. The error bars are based on a Jackknife scheme
 \citep{Efron:82,SchLoiPru:2001} using a roughly estimated correlation time of
 $50$, see Tab.~\ref{tab:corrtimes}.
}
\end{center}
\end{figure}

\subsubsection{Burning time distribution}
Another distribution of interest is the distribution of burning times,
$\PSF_{\manh}(\manh ; \theta)$. The statistics are comparatively small
for this quantity, as the burning time is defined only for the cluster
removed. However, they still seem to be good enough to allow us to make
a statement about their scaling behaviour. The rescaled data,
$\PSF_{\manh}(\manh ; \theta) \manh^{b^\ast}$ with a trial exponent
$b^\ast=1.24$ can be seen in \fref{thisto}. The intermediate part of the
distribution between $\manh=4$ and the maximum seems to bend down as
$\INFL$ increases, but the developing dip is much less pronounced than
in \fref{scaling_function}. Nevertheless, the region where a data
collapse seems possible moves out towards larger values of $\manh$,
which again prohibits simple scaling. Assuming that the bending might
become weaker for sufficiently large $\manh$ leads to a data collapse
shown in \fref{thisto_collapse}, using an exponent $\nu'=0.6$ as defined
in \Eref{def_nu}. However, only for values of $\manh \approx \manh_0$
the data possibly collapse. Again, this violates the assumption of
simple scaling, namely that there is a \emph{constant} lower cutoff
above which the behaviour is universal. 

\begin{figure}[t]
\begin{center}
\includegraphics[width=0.7\linewidth]{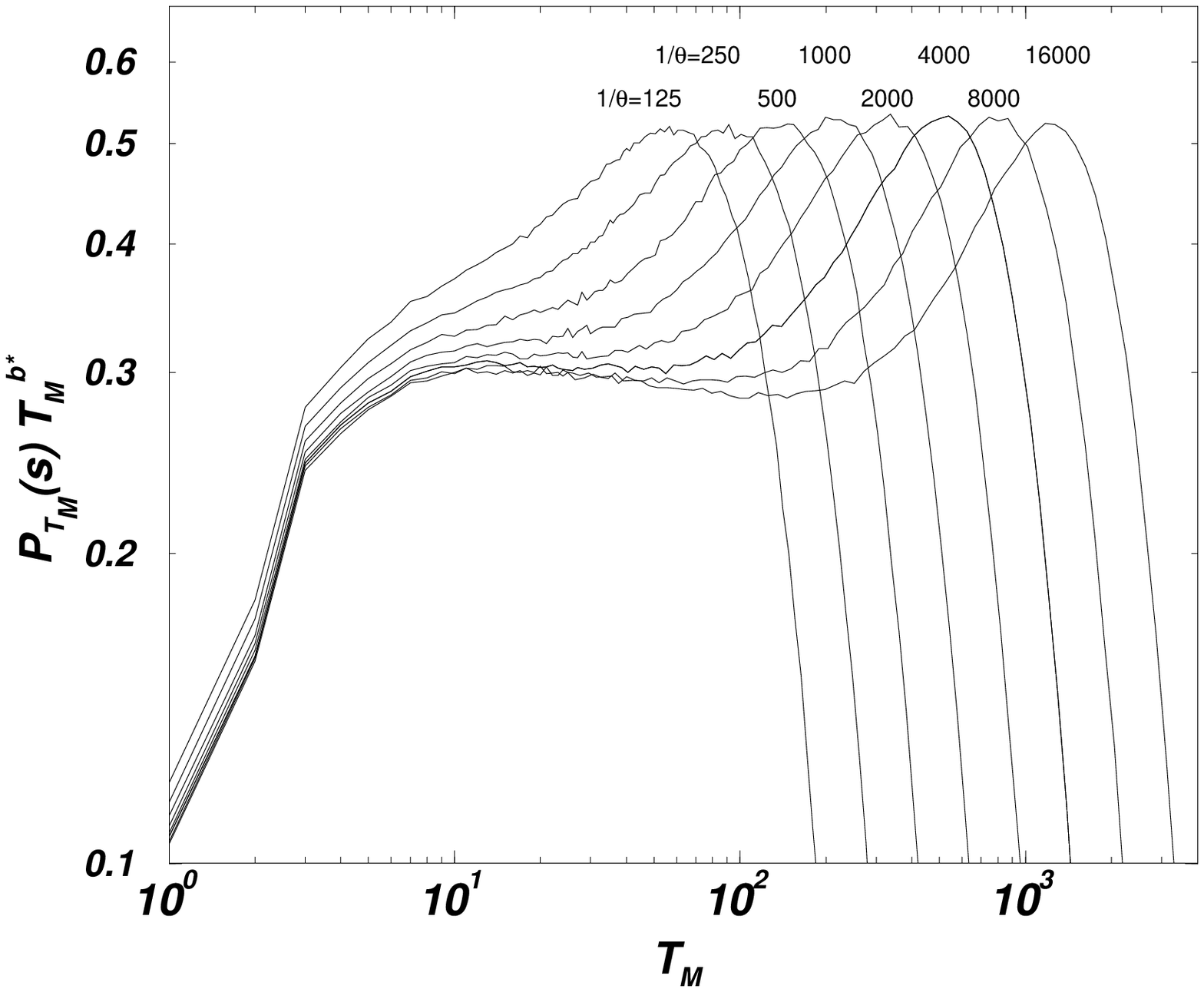}
\caption{\flabel{thisto} 
The rescaled probability distribution of the burning time,
 $\PSF_{\manh}(\manh ; \theta)$. Similar to
 \fref{scaling_function} a dip seems to form between the low
 $\manh$ region and the maximum, which again renders a data collapse
 impossible.
}
\end{center}
\end{figure}

\begin{figure}[t]
\begin{center}
\includegraphics[width=0.7\linewidth]{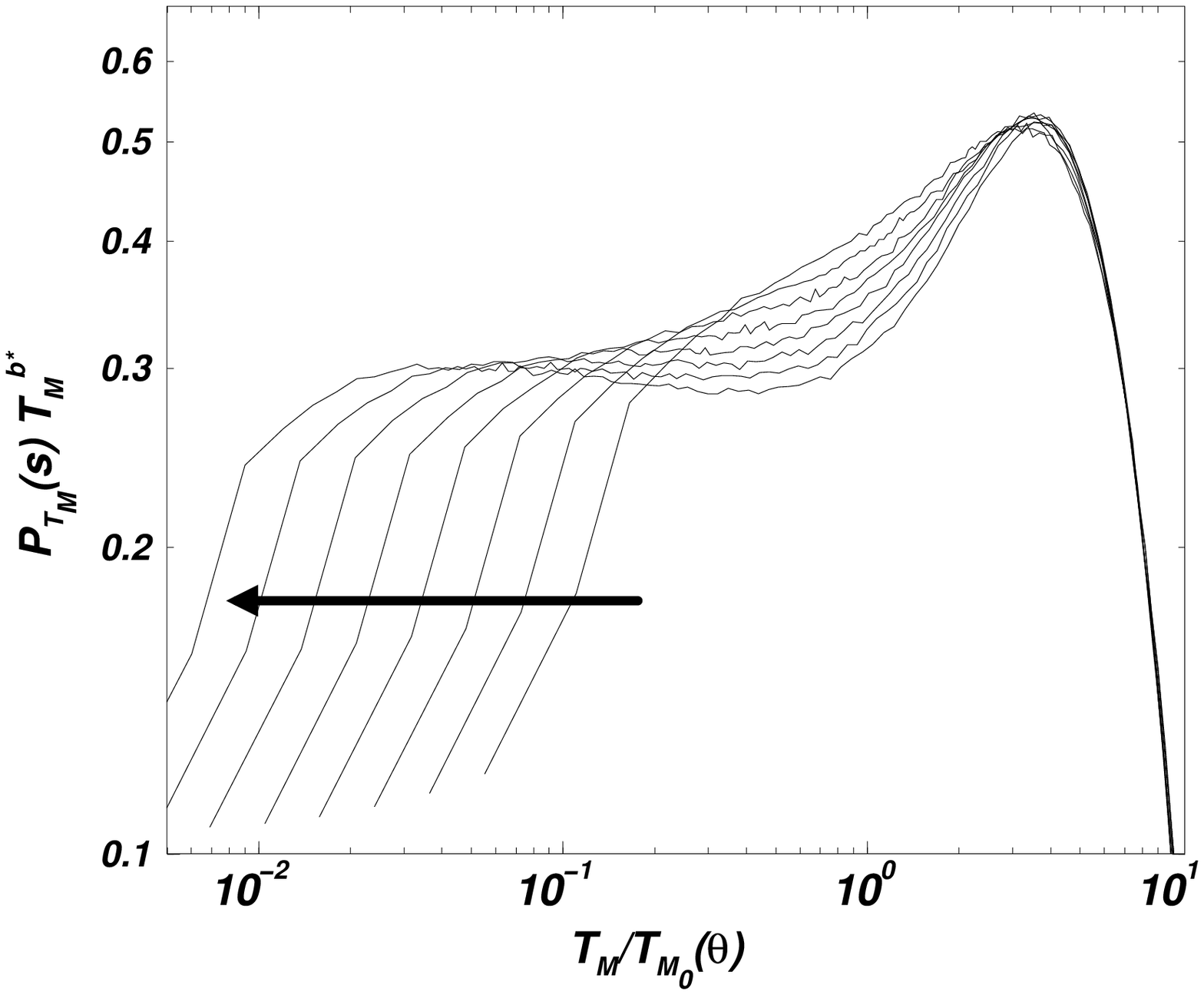}
\caption{\flabel{thisto_collapse} 
Attempt of a data collapse for $\PSF_{\manh}(\manh ; \theta)$. Only at the
 far end of the scaling function at the descent from the maximum, the
 data seem actually to collapse. This, however, is not sufficient for a
 data collapse. The big arrow points in the direction of increasing $\INFL$.
}
\end{center}
\end{figure}

\begin{figure}[ht]
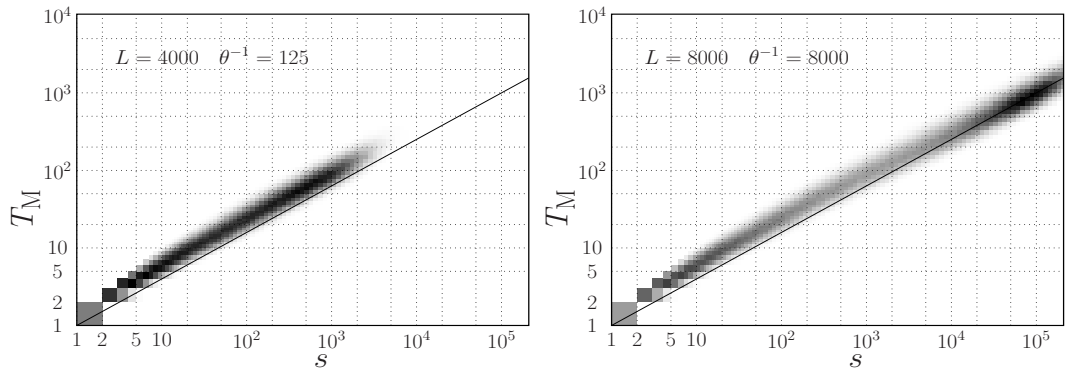

\begin{center}
\resizebox{7cm}{5cm}{
\input{figure_ST_125.tex}
}
\resizebox{7cm}{5cm}{
\input{figure_ST_8000.tex}
}
\caption{\flabel{psf}
Binned density plots of $\PSF(s, \manh ; \theta)$ for different values
 of $\theta$ on a double logarithmic scale. High densities are presented
 as dark areas. For better presentation, $\PSF(s, \manh ; \theta)$ has
 been multiplied by a factor $s^{1.7}$, tilting the distribution similar
 to those shown in \fref{scaling_function}, so that the second
 maxima in the distribution, those at large $s$ and $\manh$, are roughly
 as high as the first maxima, i.e. they show in the plot as dark as
 around $s=5$. Since $\PSF(s, \manh ; \theta)$ is a
 histogram only of burnt clusters, it contains a factor $s$ compared to
 $\dns$ (see discussion around \eref{def_rho}). Therefore, the
 exponent $2.7$ needs to be compared to $\Tast=2.10$, indicating that
 the width of $\PSF(s, \manh ; \theta)$ roughly scales like
 $s^{0.6}$, so that the reduced height of $\PSF(s, \manh ; \theta)$ is
 caused by an increase in width. This coincides well with the slope of the
 distribution, as shown by a straight line. Thus, the relative width
 remains roughly constant.  }
\end{center}
\end{figure}

The only remaining exponent of those defined in
sec.~\ref{sec:other_dists}, $\mu'$, relates the statistics of $s$ and
$\manh$. It requires the bivariate distribution $\PSF(s, \manh ;
\theta)$, as the exponent is derived from ${\mathsf E}(\manh | s)
\propto s^{1/\mu'}$, 
which is essentially equivalent to \Eref{Esmanh}. The distribution $\PSF(s, \manh ; \theta)$ is shown
in \fref{psf}. At first glance the assumption of a power law
dependence of $s$ and $\manh$ seems to be confirmed. Also the width of
the distribution seems to be very small, with almost no change over $5$
orders of magnitude in $s$. However, the plot is double logarithmic, so
that the width roughly scales like the slope, which is about $0.6$, as
shown by straight lines. This matches perfectly the exponent chosen to
rescale $\PSF$ (see caption of \Fref{psf}).

\begin{figure}[t]
\begin{center}
\includegraphics[width=0.7\linewidth]{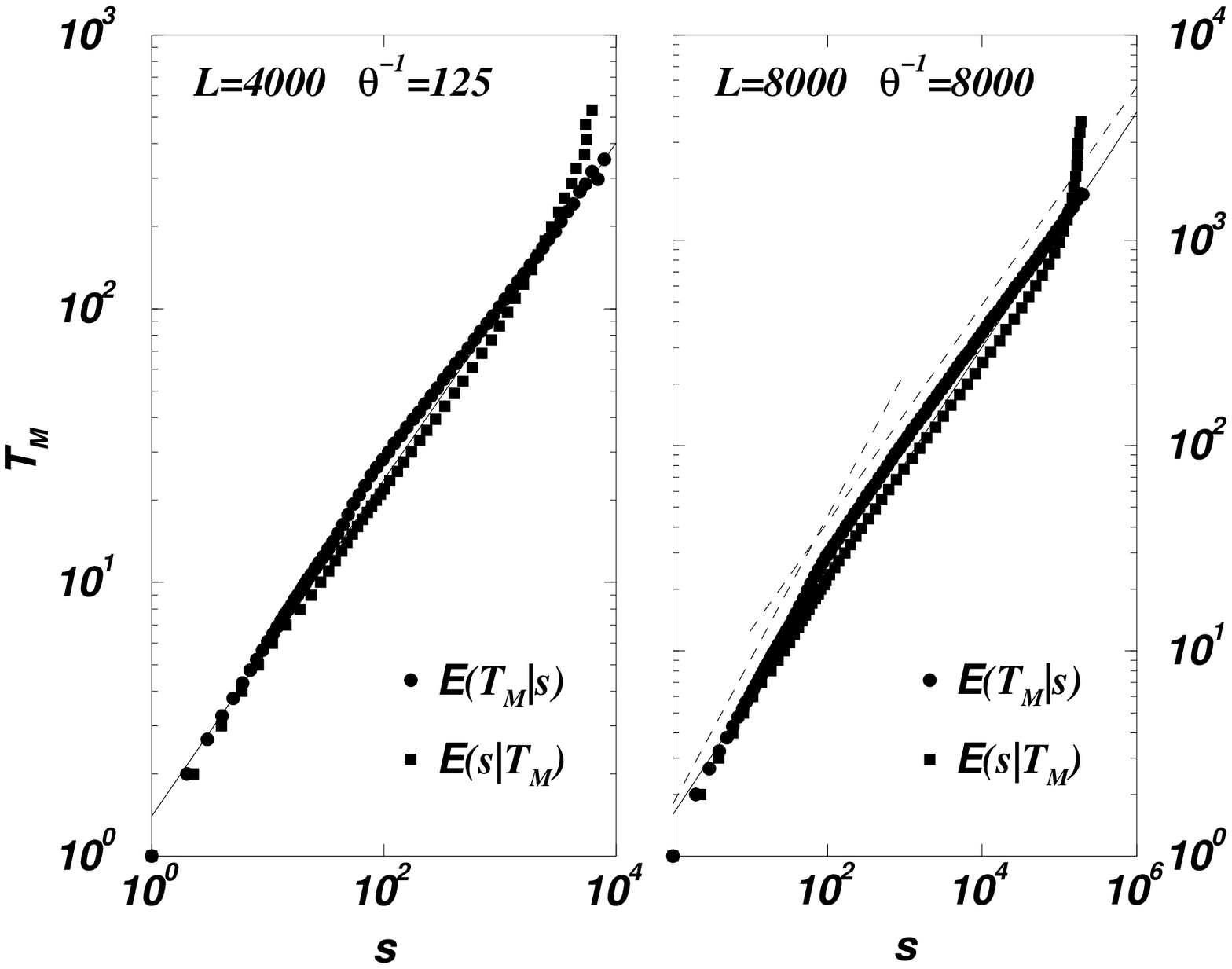}
\caption{\flabel{Esmanh} 
${\mathsf E}(\manh | s ; \theta)$ and ${\mathsf E}(\manh | s ; \theta)$,
 based on the binned histogram $\PSF(s, \manh ; \theta)$ for different
 values of $\INFL$. The straight lines in the plots are $1.4 s^{0.615}$
 for $\INFL=125$ (left hand plot) and $1.6 s^{0.57}$ for
 $\INFL=8000$. The two dashed lines in the right hand plot show alternative
 exponents $1/\mu'=0.7$ and $1/\mu'=0.53$, which are consistent with data either for
 small values of $s$ or for large values.
}
\end{center}
\end{figure}

By inspecting ${\mathsf E}(\manh | s; \theta)$ and ${\mathsf E}(s |
\manh; \theta)$  for various $\theta$, one can determine $\mu'$ as slope
in a double logarithmic plot. \fref{Esmanh} shows that $\mu'$
remains ambiguous and deviations from the expected behaviour do not
vanish as $\INFL$ is increased. Asymptotically one might expect $1/\mu' \approx 0.62$,
while $(\Tast-2)/(b^\ast-1)$ suggests $1/\mu' \approx 0.417$. The value
of $0.62$ is consistent with the rough estimate $0.6$ made in
\fref{psf}. \fref{Esmanh} also shows two other exponents, $0.53$
and $0.7$, the former being in line with the value found in literature
of $0.529(8)$ \citep{ClarDrosselSchwabl:1994}.

Conclusively it is noted that the other observable available in this
study, $\manh$, does not seem to provide an alternative way to ascribe
the DS-FFM critical behaviour in the sense of the scaling behaviour as
proposed in the literature.

\subsection{Tree density as a function of time} \label{sec:tree_density}
As mentioned above (see section~\ref{sec:finite_size_scaling}), the
density of trees, $\rhobar$, is actually a function of time. Initially, it is
periodic around the average value, with an amplitude that depends mainly
$\theta$. This amplitude decays in time and after sufficiently long
times $\rho(t)$ looks like a random
walk around $\bar{\rho}$.

\begin{figure}[t]
\begin{center}
\includegraphics[width=0.7\linewidth]{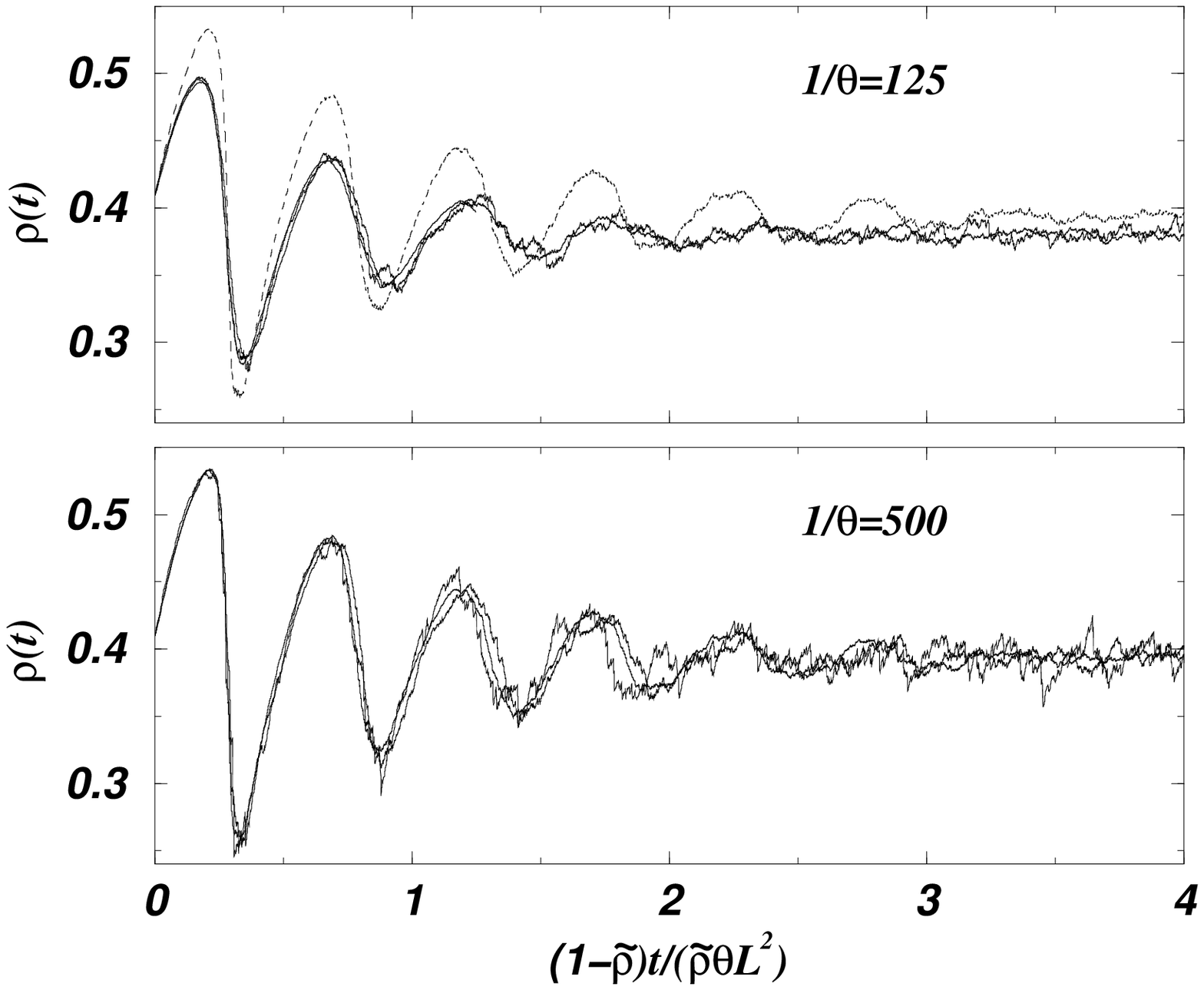}
\caption{\flabel{static_densities}
The density of trees as a function of time, plotted versus the rescaled
 time $(1-\rhobar)t/(\theta \rhobar L^2)$. Upper panel: Plot for $\INFL=125$ and
 $L=1000, 2000, 4000$ with an additional plot for $\INFL=500$ and
 $L=4000$ shown as dashed line, for comparison of period and
 amplitude. Lower panel: Same plot for $\INFL=500$ and $L=1000, 2000, 4000$.
}
\end{center}
\end{figure}

\fref{static_densities} illustrates how the period and the amplitude
depends on $\theta$ and $L$: The period is proportional to $\theta L^2$,
while the amplitude mainly depends on $\theta$, i.e. the strength of the
influx $\propto \INFL$. The reason for the
former is easy to understand: $\INFL / L^2$ is proportional to the
fraction of newly grown trees \citep{HoneckerPeschel:1997}; the change of
the tree density is roughly
\begin{equation}
 \frac{d}{dt} \rho = \frac{1-\rho}{\rho} \frac{1}{\theta L^2} - \eta(\rho(t),t)
\end{equation}
assuming that it hardly changes during the growing. Otherwise, one would
have to introduce a microscopic timescales, which 
makes it possible to measure the tree density on the timescale on which the trees are grown. The
pre-factor $(1-\rho)/\rho$ takes into account that only empty sites can
be re-occupied and that an occupied site is required for the burning to
start. The second term on the right hand side, $\eta(\rho(t),t)$, is a
noise, which represents the burning of the trees. From this equation one
can already expect that the period is roughly linear in $\theta L^2
\rhobar/(1-\rhobar)$. This has already been measured in detail by
Honecker and Peschel \citep{HoneckerPeschel:1997}; the numerical results
presented here (\Fref{static_densities}) are fully consistent with their
results.

Apart from the relevance of the periodic behaviour for the equilibration
time, the periodic behaviour of $\rho(t)$ is physically of great
significance: What distinguishes the state of the system for a given
$\rho$ at the ascending and the descending branches? Trivially, the
sequence of configurations of the system is Markovian, while the tree density alone
as a time series is certainly not. The configuration somehow manages to
``remember'' whether the tree density was increasing or decreasing
during the last update, in order to keep $\rho(t)$ periodic.

One explanation for this behaviour might be a ``growing-and-harvesting''
concept: From the initially completely random tree distribution larger
and larger patches are formed, so that larger and larger patches are
harvested by lightning. When the density reaches the maximum, for a
while the patches harvested remain large compared to the amount
grown. This is because the growing process does not actually produce
those large patches itself, but makes them available to the harvesting
by continuously connecting smaller patches in areas, where the lightning
has not yet struck. This process goes on, until almost all the trees are
newly grown, i.e. the trees are distributed almost randomly, apart from
the spatial correlation in density. The period of this process would be
proportional to the time it takes to renew the entire system, which is
$L^2 \theta \rhobar/(1-\rhobar)$, namely $L^2$ divided by $\aves{s}$,
see \eref{aves_averho}.

The time-dependent tree density gives only a hint of what actually
happens in the system. It would be very instructive to study the
two-point correlation function as a function of time to answer the
question, whether the explanation above is actually valid.

\subsection{Discussion}
From the results presented above it becomes clear that the Forest Fire Model
does not show the scaling behaviour expected for a system, which becomes
critical in the appropriate limit (namely $L\to\infty$ and $\INFL \to
\infty$). One might argue that 
another scaling ansatz could lead to a distribution which is
asymptotically scalefree in this limit, for
example a multifractal ansatz \citep{TebaldiDeMenechStella:1999} or the
one proposed in \citep{Schenk:2002}, where more than one scale is assumed
to govern the model. For an asymptotically scalefree distribution,
the scales have to diverge or to vanish in the appropriate limit. It has been
suggested already very early \citep{HoneckerPeschel:1997} that more than
one characteristic length scale can be found in the Forest Fire Model.

However, changing the scaling assumption would entail a new
\emph{definition} of the exponents $\tau$, $D$ etc., which would
therefore prohibit comparison with other results, which are based on the
assumption of simple scaling \eref{def_tau}. Moreover, introducing
multiple scales would stretch the notion of universality, especially the
universality of the scaling function, to its limits. As can be seen in
\fref{scaling_function}, the shape of the distribution function
\emph{is not universal}, i.e. the shape
of this function is unique for every single $\INFL$, even for $L\to\infty$. This is in
direct contradiction to the concept of universality, scaling and scale
invariance. 

However, it might be possible to reestablish simple scaling by
introducing another mechanism in the model, as was done for example in
the ``autoignition Forest Fire model'' \citep{Sinha-RayJensen:2000}. If
there were, for example, a mechanism parameterized by $u$, such that
\begin{equation}
 \dn(s; \theta, u) = s^\tau \GC(s/\Scutoff(\theta, u))
\end{equation}
then simple scaling might be reestablished possibly by choosing an
appropriate $u=u(\theta)$; even the cutoff, $\Scutoff$, which
were assumed to diverge with $\theta^{-1}$, would then effectively depend
only on $\theta$. Currently, there is no hint, what this new parameter
$u$ could be. 

Lise and Paczuski \citep{LisePaczuski:2001} suggested for a
similar problem in the OFC model \citep{OlamiFederChristensen:1992} to
define an exponent $\tau$ by the slope of the distribution $\PCA(s)$,
imposing the remaining background, $\mathcal{F}(s, L, \INFL)$, to be as
straight as possible: 
\begin{equation}
 \ln\left(\PCA(s)\right) = -\tau ln(s) + \mathcal{F}(s, L, \INFL)
\end{equation}
This ansatz, in fact based on a multiscaling ansatz,
would indeed allow the measurement of an exponent, however, with some
degree of ambiguity. The crucial problem with this approach is that,
firstly, it again does not allow any direct comparison to other models,
where the exponents are defined via \eref{def_tau} and that, secondly, the
notion of a presumably universal exponent hides the fact of broken
scaling.

From Section~\ref{sec:clusterdist} one might conclude that there does
not even exists a limiting distribution for $\dnst$. However, even if it
exists, that does not mean that simple scaling is obeyed and if it does,
it is still open whether the exponents are non-trivial or not and
whether the model posses any spatio-temporal correlation which do not
vanish on sufficiently large scales.

\section{Summary}
Using a new method for simulating the Forest Fire Model on large scales,
it is possible to make clear statements about the validity of the
scaling assumption of this model. The two observables investigated in
this paper suggest the model does not develop into a scale invariant state.

The method is based on the Hoshen-Kopelman algorithm
\citep{HoshenKopelman:1976} and uses a master/slave parallelisation
scheme to simulate the model on very large scales and very large sample
sizes. The key to the parallelisation is to decompose the lattice in
strips and to encode the connectivity of these strips in the border
sites. Clusters crossing these strips are then maintained by the master
node, while clusters within a strip are maintained on the local
nodes. There is almost no data exchange apart from the border
configuration, which lowers the impact on the network linking the nodes.

The resulting distribution $\PCA(s)$ is, different from other
simulations found in the literature, the distribution of \emph{all}
clusters in the system, rather than just the burnt clusters. The
resulting statistics then allows to draw clear conclusions as to what
extend the model does actually obey the scaling assumption. This turns
out not to be case. The violation of scaling is also observed in the
distribution of the burning time. Conclusively we find that there is no
reason to assume that the Drossel-Schwabl Forest Fire Model develops
into a critical state. This is in line with the conclusion by
Grassberger \citep{Grassberger:2002}, who however, still finds some signs
that the Forest Fire Model will finally show some characteristics of
standard percolation.

\begin{acknowledgments}
The authors wish to thank Andy Thomas for his fantastic technical
support. A great deal of the results in this paper was possible
only because of his work. This paper partly relies on resources
provided by the Imperial College Parallel Computing Centre. We
want to thank especially K. M. Sephton for his support.

Another part of this work was possible only because of the
generous donation made by ``I-D Media AG, Application Servers \&
Distributed Applications Architectures, Berlin''. We especially
thank M. Kaulke and O. Kilian for their support.

G.P. wishes to thank P. Grassberger, A. Honecker, I. Peschel and
K. Schenk for very helpful communication, as well as P. Anderson
and K. Dahlstedt for their advice.

The authors gratefully acknowledge the support of EPSRC.
\end{acknowledgments}

\bibliography{articles,books}
\end{document}